\documentstyle[aps,epsfig]{revtex}

\begin{document}
\title{Steady state of atoms in a resonant field with elliptical polarization}
\author{A.V.Taichenachev$^{1}$, A.M.Tumaikin$^{2}$, V.I.Yudin$^{2}$, and G.Nienhuis$%
^{3}$}
\address{(1) Novosibirsk State University,\\
Pirogova 2, Novosibirsk 630090, Russia}
\address{(2) Institute of Laser Physics SD RAS, Lavrent'eva 13/1,\\
Novosibirsk 630090, Russia}
\address{(3) Huygens Laboratorium, University of Leiden, \\
P. O. Box 9504, 2300 RA Leiden, The Netherlands}
\maketitle

\begin{abstract}
We present a complete set of analytical and invariant expressions for the
steady-state density matrix of atoms in a resonant radiation field with
arbitrary intensity and polarization. The field drives the closed dipole
transition with arbitrary values of the angular momenta $J_{g}$ and $J_{e}$
of the ground and excited state. The steady-state density matrix is
expressed in terms of spherical harmonics of a complex direction given by
the field polarization vector. The generalization to the case of broad-band
radiation is given. We indicate various applications of these results.
\end{abstract}


\section{Introduction}

An atomic medium driven by a resonant light field represents a prototype
problem in atomic physics and non-linear optics. At low density the effects
of the atom-atom interaction are small, and the central remaining problem is
specified by a single atom in a resonant radiation field. As is well-known,
the basic processes are the absorption and emission of photons. The three
universal conservation laws (energy, linear momentum and angular momentum)
correspond to three different aspects in these processes. The main exchange
of energy between atom and field corresponds to the radiative transition
between the atomic energy levels. The momentum exchange gives rise to recoil
of the atom, which is the basis of the mechanical action of light. The main
angular momentum exchange arises from the photon spin. Its proper
description requires consideration of the light polarization, and the
degeneracy of atomic energy levels. Obviously, these processes occur
simultaneously, and in a correlated fashion. The recoil effect is usually
small, due to the smallness of the photon momentum ($\hbar k$) as compared
to typical values of the atom momentum ($M \bar{v}$). In contrast, the
(spin) angular momentum of photons $\hbar$ is of the same order of magnitude
as the internal angular momentum of the atomic states.

A large fraction of theoretical studies of atoms in radiation fields only
considers non-degenerate energy eigenstates, without taking into account the
magnetic degeneracy of energy levels. In the sense of the group of space
rotations, this approach corresponds to a scalar model of the atom, which
accounts for exchange of energy and momentum, but not of angular momentum.
This model allows to understand many processes arising from the resonant
interaction of atoms with radiation \cite
{eberly78,let&cheb77,rautian79,minogin87,kazantsev91}. However, effects of
polarization in combination with the magnetic degeneracy of atomic levels
cannot be ignored in many cases. The problem remains reasonably tractable in
the special cases of linear or circular polarization. In these cases, there
is an obvious choice of the quantization axis, so that the submatrices of
the density matrix for the ground and the excited level remain diagonal at
all times. For arbitrary elliptical polarization the situation is
appreciably more complex. There are various experimental situations where
elliptical polarization is quite essential. An important example is the
phenomenon of coherent population trapping (CPT) of atoms driven by an
elliptically polarized light field \cite{smirnov89}. Another example is
cooling and trapping of atoms in light fields with polarization gradients
\cite{lasercooling}, where the strong correlation between the processes of
linear and angular momentum exchange can lead to atom temperatures down to
the single-photon recoil energy ($\sim 10^{-6}K$). Here continuous spatial
variations of the polarization are crucial, which, except for special cases,
give rise to elliptical polarization. Models with non-degenerate atomic
states can only describe the Doppler limit of laser cooling ($\sim 10^{-3}K$%
) \cite{kazantsev91}.

A central part of these processes is the resonant interaction of an
elliptically polarized light field with a closed atomic transition between
degenerate energy levels. In this case the light-induced anisotropy of the
atomic state is long-lived. This enables one to accumulate information on
very weak couplings, which allows for high-resolution spectroscopy. As noted
above, for many cases one can consider the recoil effect as a small
perturbation of the order of $\hbar k/M\bar{v}\ll 1$. In zeroth order the
atom has a constant linear momentum. In this case only the exchange of
energy and angular momentum between atom and field is taken into account,
and the state of the atom is fully described by the density matrix for the
internal states. The remaining field effects (light shifts, field
broadening, change of population and coherence etc.) are caused by the
stimulated and spontaneous transitions, which are described by the
generalized optical Bloch equations (GOBE) for the internal atomic density
matrix. Depending on the light intensity and interaction time, three
limiting cases can be distinguished. The first case occurs for short light
pulses, when the interaction time is so short that relaxation processes can
be neglected. Then the interaction of atoms with the field has a coherent
character, and it can be described by the time-dependent Schr\"{o}dinger
equation for the atomic wavefunction. Typical effects are coherent transient
processes, such as Rabi oscillations or photon echoes, which have been
thoroughly analyzed \cite{eberly78,manikin84}, with or without inclusion of
the magnetic degeneracy of atomic states \cite{PPE}. The second case occurs
when the interaction time is long compared with the spontaneous lifetime,
but still short compared with the absorption time, which determines the rate
of optical pumping. In this case one can use perturbation theory. This
situation arises for laser beams of low intensity and small diameter, when
time-of-flight effects become important. Solutions of the optical Bloch
equations corresponding to this case have been obtained with the use of
irreducible-tensor techniques for arbitrary Zeeman and hyperfine level
structures \cite{velich89}. Finally, when the interaction time is so long
that perturbation theory becomes inapplicable, one has to find the
steady-state solution of the GOBE. This situation appears either for slow
atoms interacting with light, such as in optical molasses, or in the case of
light beams with high intensity or large diameter. A common restriction is
that only a closed transition between two degenerate atomic levels is
considered, so that the total population is conserved.

The GOBE are an essential ingredient of the description of sub-Doppler laser
cooling by fields with polarization gradients \cite{Dalibard}. We here
recall just a few representative cases. In the semiclassical theory of laser
cooling authors concentrated their efforts on the velocity-dependent
steady-state density matrix. The presented analytical results were, however,
restricted to transitions with specific small values of the angular momenta
of the states \cite{sctheory}. Berman et al. \cite{berman91} have formulated
the GOBE for arbitrary field polarization and arbitrary atomic energy level
structure, using the irreducible tensor representation. They demonstrate
that the sub-Doppler light forces and the sub-natural resonances in
non-linear spectroscopy are closely related. Similar results have been
obtained in \cite{nienhuis91}, where a general relationship between the
light force and the non-linear polarizability tensor has been derived.

For linear or circular polarization, the steady-state solutions of the GOBE
have been discussed in various papers \cite{macek74,nienhuis82,gao93} for
arbitrary values of $J_{g}$ and $J_{e}$. For arbitrary polarization, the
symmetry is reduced, and the steady state for arbitrary $J_{g}$ and $J_{e}$
represents a complex mathematical problem. The number of equations, which is
equal to the number of elements of the density matrix, amounts to $%
4(J_{g}+J_{e}+1)^{2}$. The steady-state density matrix was found in
analytical form only for transitions involving specific small values of the
angular momentum ($J_{g}=0,\,1/2,\,1$) \cite
{dalibard84,dalibard89,suter91,davis92,molmer94}. Recently we have discussed
the steady state of atoms in light fields with arbitrary polarization, for
transitions with $J_{e}-J_{g}=0$ \cite{taich95,taich99epl} or $1$ \cite
{nienhuis98epl}. In these cases, the structure of the solutions looked
remarkably different. An invariant approach to the general problem, based on
the expansion of the density matrix in bipolar harmonics of complex
directions was developed in \cite{bezverbny00}. Nasyrov \cite{nasyrov01} has
suggested an alternative approach to the steady-state density matrix, using
the semiclassical Wigner representation of angular-momentum orientation .
This method seems especially useful at large angular momentum $J\gg 1$. The
results are in good qualitative agreement with our exact solution.

In the present paper we give unified exact analytical expressions for the
steady state of atoms driven by light with an arbitrary polarization, for
all possible dipole transitions. Radiative relaxation is included in the
description, and the results are presented in an invariant form. The
analysis is based upon the group-theoretical properties of the transition
dipole. The polarization direction in real space is reflected in the density
matrix for the two subspaces corresponding to the ground and excited state.
The symmetry group for the problem is the group of rotations $SU(2)$, which
is an important leading principle for the search of the solution, as well as
for its presentation in invariant form. The general discussion allows us to
clarify the similarities in the different cases, which were not obvious at
all in the separate treatments.

The nature of the steady state strongly depends on the value of $J_e - J_g$.
For dipole-allowed transitions, this difference can be $-1$, $0$ or $1$. For
the case that $J_e = J_g = J$ we should moreover distinguish the cases that $%
J$ is integer or half-integer. This leads to four classes of transitions.

\begin{itemize}
\item[a)]  Transitions $J_{g}=J\rightarrow J_{e}=J-1$. In this case atoms
are optically pumped into dark states, where they do not couple to the light
field. This is the phenomenon of coherent population trapping. These dark
states are linear superpositions of the Zeeman substates of the ground
level, defined as eigenstates of the resonant interaction Hamiltonian with
zero eigenvalue. Hence, there are no light shifts. For the present class of
transitions, there are two independent dark states, which span a
two-dimensional space. This space depends only on the polarization, and it
is independent of the field intensity and the detuning. Both the atomic
dynamics in the field and the steady state depend on the initial state.

\item[b)]  Transitions $J_{g}=J\rightarrow J_{e}=J$ with $J$ integer. In
this case a single dark state exists, and CPT takes place, so that in the
steady state the atom is in this unique pure state. Therefore, the steady
state does not depend on the initial conditions, the intensity or the
detuning.

\item[c)]  Transitions $J_{g}=J\rightarrow J_{e}=J$ with $J$ half-integer.
For this class no dark state exists, and CPT does not occur. The only
exception is the case of circular polarization, where a single dark state
does occur. The steady state is uniquely defined, but now it depends both on
the polarization, the intensity and the detuning. Moreover, it is not a pure
state. In fact, it has the remarkable property that the excited-state
submatrix of the density matrix is fully isotropic, which makes the
analytical expression for the entire density matrix particularly simple.

\item[d)]  Transitions $J_{g}=J\rightarrow J_{e}=J+1$. For this class of
transitions the steady-state solution is unique. There is no dark state. The
excited-state submatrix is always anisotropic.
\end{itemize}

Only in the cases c) and d) does a steady-state excitation exist. In both
cases, the anisotropy of the excited state and of the optical coherences
depends only on the polarization of the driving field. The intensity and the
detuning enter only as an overall multiplicative factor. Moreover, in both
cases we find that the submatrices both for the excited and the ground state
are even functions of the detuning. In the cases a) and b) only the ground
state is populated in the steady state, and the excited-state submatrix and
the optical coherences vanish. The occurrence of dark states and
velocity-selective CPT allows to reach cooling below the recoil limit \cite
{Aspect}, and it has been studied by many authors \cite{vscpt}. In the case
of $J_{g}=1$ an invariant form of the dark states in an elegant vector
notation has been used for the analysis of CPT in 2D and 3D \cite{invvscpt}.
Here we extend an invariant approach to all the dark transitions.

The remainder of the paper is organized as follows. First we discuss the
general structure of the generalized optical Bloch equations (Sec. II),
which leads in Sec. III to the definition of a natural basis of states that
depend only on the light polarization. We separately discuss the cases with
(Sec. IV) or without (Sec. V) dark states. In the latter case, we discuss
the effect of optical pumping on the degree of excitation and the AC Stark
shift in Sec. VI. The generalization to the situation of broad-band
radiation is given in Sec. VII.

\section{Formulation of the problem}

We consider a closed atomic transition $J_{g}\rightarrow J_{e}$ with a
transition frequency $\omega _{0}$ of an atom at a given position. In the
present paper we will not consider the translational motion of the atom.
This corresponds either to the case of very slow atoms or to the case of a
travelling plane wave, where the atomic motion at a given velocity leads
only to a Doppler frequency shift. The transition is driven by a
monochromatic radiation field with frequency $\omega $ and arbitrary
polarization ${\bf e}$. The time-dependent electric field vector at the
position of the atom is given by
\begin{equation}
{\bf E}=E_{0}{\bf e}\exp (-i\omega t)+c.c.  \label{field}
\end{equation}
where the polarization vector is
\begin{equation}
{\bf e}=\sum_{q=0,\pm 1}e_{q}{\bf e}_{q}^{*}\;=\sum_{q=0,\pm 1}(-1)^{q}e_{-q}%
{\bf e}_{q}.  \label{polarization}
\end{equation}
Here $E_{0}$ is the complex field amplitude, $e_{q}={\bf e}\cdot {\bf e}_{q}$
is the covariant spherical component of the polarization vector ${\bf e}$,
and the spherical basis vectors of polarization are defined by $\{{\bf e}%
_{0}={\bf e}_{z};\;{\bf e}_{\pm 1}={\mp }({\bf e}_{x}\pm i{\bf e}_{y})/\sqrt{%
2}\}$. Notice that ${\bf e}_{q}^{*}=(-1)^{q}{\bf e}_{-q}$. The vector ${\bf e%
}$ is normalized, so that ${\bf e}^{*}\cdot {\bf e}=1$ and without loss of
generality we assume that its real and imaginary parts are orthogonal, which
implies that $\mbox{Im}({\bf e}\cdot {\bf e})=0$. Then the two vectors $%
\mbox{Re}\;{\bf e}$ and $\mbox{Im}\;{\bf e}$ are the axes of the
polarization ellipse.

It is always possible to use a coordinate frame where only two of the
components $e_{q}$ are nonzero. There are two possibilities. The
conventional choice is that the $Oz$ axis is chosen normal to the
polarization plane. In this coordinate system the vector ${\bf e}$ is the
sum of the two opposite circular unit vectors ${\bf e}_{\pm 1}$. If the $Ox$
axis is directed along the major semiaxis of the polarization ellipse (see
Fig. 1a), ${\bf e}$ is written as
\[
{\bf e}={\bf e}_{x}\;\cos {\varepsilon }+i{\bf e}_{y}\;\sin {\varepsilon }=-%
{\bf e}_{+1}\;\sin {(\varepsilon +\pi /4)}+{\bf e}_{-1}\cos {(\varepsilon
+\pi /4)}\,,
\]
where the ellipticity angle $\varepsilon $ can take the values $-\pi /4\le
\varepsilon \le \pi /4$. Obviously, $|\tan \,\varepsilon |$ is equal to the
ratio of the minor semiaxis to the major semiaxis and the sign of $%
\varepsilon $ determines the helicity.

Another choice is called the natural coordinate frame, which was introduced
in \cite{tumaikin90}. When we represent the polarization ellipse as the
intersection of a cylinder with a plane, the natural frame $Ox^{\prime
}y^{\prime }z^{\prime }$ is specified by the requirement that the axis $%
Oz^{\prime }$ is the axis of the cylinder, while the axis $Oy^{\prime }$
coincides with the axis $Oy$. The minor semiaxis of the ellipse coincides
with the radius of the cylinder (Fig. 2a). Then the polarization ${\bf e}$
is the superposition of a linear component along $Oz^{\prime }$, and one
circular component. The two frames are connected by a rotation along the
axis $Oy$ over an angle $\theta $ obeying the relation
\[
\cos {\theta }=|\tan \,{\varepsilon }|\,.
\]
In the natural frame, the polarization vector is specified as
\begin{equation}
{\bf e}={\bf e}_{0}^{\prime }\sqrt{\cos (2\varepsilon )}-{\bf e}_{\pm
1}^{\prime }\sqrt{2}\sin (\varepsilon ),  \label{conf2}
\end{equation}
where the helicity of the spherical unit vector ${\bf e}_{\pm 1}^{\prime }$
corresponds to the sign of $\varepsilon $. In general there are two possible
choices for the cylinder, corresponding to opposite signs of the rotation
angle $\theta $. Notice that when the polarization ${\bf e}$ is represented
in terms of the Stokes vector as a point on the Poincar\'{e} sphere, the
angle $\theta $ is equal to the polar angle of this point \cite{born}.

The quantum kinetic equation for the density matrix $\widehat{\rho }$ of the
internal state of the atom in the external field (\ref{field}) has the form:
\begin{equation}
\frac{\partial }{\partial t}\widehat{\rho }=-\frac{i}{\hbar }\left[ \widehat{%
H}_{0},\widehat{\rho }\right] -\frac{i}{\hbar }\left[ -\widehat{{\bf d}}%
\cdot {\bf E}(t),\widehat{\rho }\right] -\widehat{\Gamma }\{\widehat{\rho }%
\}\;.  \label{general}
\end{equation}
Here $\widehat{H}_{0}$ is the Hamiltonian describing the energy of the two
resonant levels of the free atom and $\widehat{{\bf d}}$ is the dipole
operator connecting the two levels. The radiative relaxation is described by
the operator $\widehat{\Gamma }\{\widehat{\rho }\}$. All operators are
represented as matrices on the Zeeman basis of the ground and excited
levels, with states $\{|J_{g},\mu _{g}\rangle \}$ and excited $\{|J_{e},\mu
_{e}\rangle \}$. The density matrix $\widehat{\rho }$ can be separated in
four matrix blocks, where the matrices $\widehat{\rho }_{gg}$ and $\widehat{%
\rho }_{ee}$ are the submatrices for the ground and excited state, and the
off-diagonal blocks $\widehat{\rho }_{eg}$ and $\widehat{\rho }_{ge}$
describe the optical coherences. In the rotating-wave approximation the time
dependence of the kinetic equation can be removed by introducing the
transformed optical coherences as
\begin{equation}
\widehat{\rho }_{eg}=\exp (-i\omega t)\widehat{\overline{\rho }}_{eg}\;\;\;\;%
\widehat{\rho }_{ge}=\exp (i\omega t)\widehat{\overline{\rho }}_{ge}.
\label{average}
\end{equation}
The resulting system of generalized optical Bloch equations (GOBE) can be
expressed in the dimensionless dipole operator $\widehat{{\bf D}}$, which
couples the ground state to the excited state. It is specified by the
definition of its spherical components as
\begin{equation}
\widehat{D}_{q}\equiv \widehat{{\bf D}}\cdot {\bf e}_{q}=\sum_{(\mu
)}|J_{e},\mu _{e}\rangle \,C_{J_{g}\,\mu _{g}\;1\,q}^{J_{e}\,\mu
_{e}}\,\langle J_{g},\mu _{g}|\,,  \label{Qq}
\end{equation}
so that their matrix elements are equal to the Clebsch-Gordan coefficients $%
C_{J_{g}\,\mu _{g}\;1\,q}^{J_{e}\,\mu _{e}}$. Stimulated transitions are
described by the operator $\widehat{V}$, which is the component of the
vector operator $\widehat{{\bf D}}$ in the polarization direction:
\begin{equation}
\widehat{V}=\widehat{{\bf D}}\cdot {\bf e}\,.  \label{VQe}
\end{equation}
Then the GOBE take the form
\begin{equation}
\left( \frac{\partial }{\partial t}+\frac{\gamma }{2}-i\delta \right)
\widehat{\overline{\rho }}_{eg}=i\Omega \left[ \widehat{V}\widehat{\rho }%
_{gg}-\widehat{\rho }_{ee}\widehat{V}\right]   \label{gobe1}
\end{equation}
\begin{equation}
\left( \frac{\partial }{\partial t}+\frac{\gamma }{2}+i\delta \right)
\widehat{\overline{\rho }}_{ge}=i\Omega ^{*}\left[ \widehat{V}^{\dagger }%
\widehat{\rho }_{ee}-\widehat{\rho }_{gg}\widehat{V}^{\dagger }\right]
\label{gobe2}
\end{equation}
\begin{equation}
\left( \frac{\partial }{\partial t}+\gamma \right) \widehat{\rho }%
_{ee}=i\left[ \Omega \widehat{V}\widehat{\overline{\rho }}_{ge}-\Omega ^{*}%
\widehat{\overline{\rho }}_{eg}\widehat{V}^{\dagger }\right]   \label{gobe3}
\end{equation}
\begin{equation}
\frac{\partial }{\partial t}\widehat{\rho }_{gg}-\gamma \sum_{q=0,\pm 1}%
\widehat{D}_{q}^{\dagger }\,\widehat{\rho }_{ee}\,\widehat{D}_{q}=i\left[
\Omega ^{*}\widehat{V}^{\dagger }\widehat{\overline{\rho }}_{eg}-\Omega
\widehat{\overline{\rho }}_{ge}\widehat{V}\right] ,  \label{gobe4}
\end{equation}
with the normalization

\begin{equation}
{\rm Tr}\{\widehat{\rho }_{gg}\}+{\rm Tr}\{\widehat{\rho }_{ee}\}=1.
\label{trace}
\end{equation}
Here $\delta =\omega -\omega _{eg}$ is the detuning, $\omega
_{eg}=(E_{e}-E_{g})/\hbar $ is the transition frequency, $\gamma $ is the
radiation relaxation rate and $\Omega =E_{0}\langle J_{e}||d||J_{g}\rangle
/\hbar $ is the generalized Rabi frequency, expressed in the complex field
amplitude $E_{0}$ and the reduced dipole matrix element $\langle
J_{e}||d||J_{g}\rangle $ that determines the strength of the transition. The
summation in eq. (\ref{gobe4}), which describes the feeding of the ground
state by spontaneous decay, runs over the three possible independent
polarizations $\{{\bf e}_{0},{\bf e}_{\pm 1}\}$ of spontaneous emission.
Conservation of the total population of the closed transition is ensured by
the relation
\begin{equation}
\sum_{q=0,\pm 1}\widehat{D}_{q}\,\widehat{D}_{q}^{\dagger }=\widehat{\Pi }%
_{e}\;.  \label{norm}
\end{equation}
When acting on an isotropic excited state, this feeding term is proportional
to
\begin{equation}
\sum_{q=0,\pm 1}\widehat{D}_{q}^{\dagger }\,\widehat{D}_{q}=\frac{2J_{e}+1}{%
2J_{g}+1}\,\widehat{\Pi }_{g}\,.  \label{isotrop}
\end{equation}
We introduced $\widehat{\Pi }_{g}$ and $\widehat{\Pi }_{e}$ as the
projectors on the ground state and the excited state. The dynamical
equations (\ref{gobe1})-(\ref{gobe4}) represent the generalized optical
Bloch equations, which describe transient processes as generalized damped
Rabi oscillations, optical nutation, free induction decay, etc., as well as
optical pumping effects, that lead to an anisotropic distribution of atoms
over the magnetic sublevels.

Equations for the steady state are obtained by setting all time derivatives
to zero in eqs. (\ref{gobe1})-(\ref{gobe4}), which gives a set of linear
equations for the density-matrix elements. By using eqs. (\ref{gobe1}) and (%
\ref{gobe2}), the steady-state optical coherences can be directly expressed
in the population submatrices $\widehat{\rho }_{ee}$ and $\widehat{\rho }%
_{gg}$ as
\[
\widehat{\overline{\rho }}_{eg}=\frac{-i\Omega }{\gamma /2-i\delta }\left[
\widehat{V}\widehat{\rho }_{gg}-\widehat{\rho }_{ee}\widehat{V}\right] \;,
\]
\begin{equation}
\widehat{\overline{\rho }}_{ge}=\frac{-i\Omega ^{*}}{\gamma /2+i\delta }%
\left[ \widehat{V}^{\dagger }\widehat{\rho }_{ee}-\widehat{\rho }_{gg}%
\widehat{V}^{\dagger }\right] .  \label{current}
\end{equation}
After substitution in (\ref{gobe3}) and (\ref{gobe4}), this leads to closed
equations for the population submatrices in the form
\begin{equation}
\gamma \widehat{\rho }_{ee}=-\gamma S/2\{\widehat{V}\widehat{V}^{\dagger },%
\widehat{\rho }_{ee}\}+\gamma S\widehat{V}\widehat{\rho }_{gg}\widehat{V}%
^{\dagger }+i\delta S[\widehat{V}\widehat{V}^{\dagger },\widehat{\rho }%
_{ee}]\;,  \label{pope}
\end{equation}
\begin{equation}
-\gamma \sum_{q=0,\pm 1}\widehat{D}_{q}^{\dagger }\widehat{\rho }_{ee}%
\widehat{D}_{q}=-\gamma S/2\{\widehat{V}^{\dagger }\widehat{V},\widehat{\rho
}_{gg}\}+\gamma S\widehat{V}^{\dagger }\widehat{\rho }_{ee}\widehat{V}%
-i\delta S[\widehat{V}^{\dagger }\widehat{V},\widehat{\rho }_{gg}];
\label{popg}
\end{equation}
where $\{\,,\,\}$, $[\,,\,]$ indicate an anticommutator and a commutator,
respectively, and where
\begin{equation}
S=\frac{|\Omega |^{2}}{\gamma ^{2}/4+\delta ^{2}}  \label{sat}
\end{equation}
is the saturation parameter, which is proportional to the light intensity
and to the global oscillator strength of the transition. The left-hand sides
of eqs. (\ref{pope}) and (\ref{popg}) describe spontaneous processes, {\em %
i.e.} the radiative damping of the excited level and the spontaneous
transfer of population and Zeeman coherence from the excited to the ground
level. The terms in the right-hand sides that are proportional to the
optical pumping rate $\gamma S$ represent light-induced loss (with the minus
sign) and gain (with the plus sign) of the levels. The commutator terms in
the right-hand sides, proportional to $\delta S$, describe the AC Stark
effect. They contain the light-shift operators in the ground and excited
level
\begin{equation}
\widehat{{\cal E}}_{g}=\delta S\widehat{V}^{\dagger }\widehat{V}\;,\;%
\widehat{{\cal E}}_{e}=-\delta S\widehat{V}\widehat{V}^{\dagger }\,,
\label{Eg}
\end{equation}
which play the role of an effective Hamiltonian for the two levels.

The steady-state solution of the GOBE corresponds to the limit $t\to \infty $%
, with $t$ the interaction time. In practice this means that $t$ is larger
than the largest relaxation time in the internal degrees of freedom. In the
case of a degenerate ground state at low saturation this largest time is of
the order of $(\gamma S)^{-1}$, which is the inverse of the rate of optical
orientation in the ground state. For large saturations $S>1$ the largest
relaxation time is the excited-state lifetime $\gamma ^{-1}$. Thus, the
conditions for the steady-state regime can be written as
\begin{equation}
{\rm min}\{\gamma t,\gamma St\} \gg 1\,.  \label{ssregime}
\end{equation}

\section{Natural basis of states}

\subsection{Commutation of density matrix and light-shift operators}

In the special case of linear or circular polarization, the Zeeman substates
$|J_g,\mu_g \rangle$ and $|J_e,\mu_e \rangle$ constitute an obvious natural
basis of substates in which to express the density matrix. For linear
polarization, one chooses the quantization axis parallel to the polarization
direction, so that the polarization vector ${\bf e}$ is equal to the
spherical unit vector ${\bf e}_q$ with $q=0$. For circular polarization, the
polarization vector ${\bf e}$ is equal to the spherical unit vector ${\bf e}%
_q$ with $q= \pm 1$, provided that the quantization axis is chosen normal to
the polarization plane. In both cases, the operator $\widehat{V}$ couples
each substate $|J_g,\mu_g \rangle$ to a single excited state $|J_e,\mu_e
\rangle$, which $\mu_e = \mu_g$ (linear polarization) or $\mu_e = \mu_g \pm
1 $ (circular polarization). In this case it can be easily checked from eqs.
(\ref{gobe1})-(\ref{gobe4}) that the equations for the populations do not
mix with those for the Zeeman coherences. Also, eqs. (\ref{pope}) and (\ref
{popg}) show that the steady-state solutions $\widehat{\rho}_{ee}$ and $%
\widehat{\rho}_{gg}$ are diagonal on the basis of the Zeeman substates $%
|J,\mu \rangle $. Since also the light-shift operators $\widehat{{\cal E}}_g$
and $\widehat{{\cal E}}_e$ are diagonal on the Zeeman substates, this
implies that the steady-state density matrices $\widehat{\rho}_{gg}$ and $%
\widehat{\rho}_{ee}$ commute with the light-shift operators $\widehat{{\cal E%
}}_g$ and $\widehat{{\cal E}}_e$, so that for linear or circular
polarization we find
\begin{equation}  \label{statementG}
[\widehat{\rho}_{gg},\widehat{{\cal E}}_g]=0 \;,\;\;\; [\widehat{\rho}_{ee},%
\widehat{{\cal E}}_e]=0 \;.
\end{equation}
Moreover, when all Zeeman coherences are zero initially at $t=t_0$, they
remain zero for all later times $t>t_0$.

One might be tempted to believe that also for arbitrary elliptical
polarization a basis of states can be chosen for which populations and
coherences do not mix. However, this is not true. It has been shown in ref.
\cite{nienhuis86} that spontaneous decay can create coherence between
eigenstates of $\widehat{{\cal E}}_g$, even if they do not exist initially.
In general the time-dependent solutions $\widehat{\rho}_{gg}$ and $\widehat{%
\rho}_{ee}$ for arbitrary polarization will not commute with the light-shift
operators at all times.

Nevertheless, in this paper we shall prove that for the steady-state
solutions for all classes of transitions the commutation rules (\ref
{statementG}) are valid for arbitrary elliptical polarization and for all
classes of transitions. The proof is rather different for the various
classes, so that it is most convenient to give the proof while discussing
the expression for the steady state for each class separately. An immediate
consequence of the commutation rules (\ref{statementG}) is that the
steady-state density matrix is diagonal in the eigenstates of the
light-shift operators. This implies also that the last terms in eqs. (\ref
{pope}) and (\ref{popg}) vanish, so that the steady-state population
submatrices depend on the detuning $\delta$ and the spontaneous-decay rate $%
\gamma$ only through the saturation parameter $S$.

\subsection{Eigenbasis of light-shift operators}

We are interested in the eigenstates of the operators $\widehat{V}^{\dagger}%
\widehat{V}$ and $\widehat{V}\widehat{V}^{\dagger}$. The corresponding
eigenvalues are real and non-negative, so that we can write
\begin{equation}  \label{diagV2}
\widehat{V}^{\dagger}\widehat{V}|(g)i\rangle=\lambda_i^2|(g)i\rangle \,;\;\;
\widehat{V}\widehat{V}^{\dagger}|(e)j\rangle=\lambda_j^2|(e)j\rangle
\end{equation}
with $\lambda_i$ real, and with eigenstates $|(g)i\rangle$ in the ground and
$|(e)j\rangle$ in the excited level. For given values of $J_e$ and $J_g$,
the eigenstates and eigenvalues are fully determined by the polarization of
the driving field, and they do not depend on the detuning or the intensity.
The states $|(e)j\rangle$ and $|(g)i\rangle$ form the natural bases for the
excited and the ground level. The operators $\widehat{{\cal E}}_g$ and $%
\widehat{{\cal E}}_e$ are diagonal with diagonal elements $\delta S
\lambda_i^2$ and $- \delta S \lambda_j^2$. Operating with the coupling
matrices $\widehat{V}$ and $\widehat{V}^{\dagger}$ on the first equation (%
\ref{diagV2}) shows that $\widehat{V} |(g)i\rangle$ is eigenstate of $%
\widehat{V}\widehat{V}^{\dagger}$ with eigenvalue $\lambda_i^2$. Hence we
may assume that $\widehat{V} |(g)i\rangle$ is proportional to $|(e)i\rangle$%
. A proper choice of the phases of the eigenstates $|(e)i\rangle$ then leads
to the expression
\begin{equation}  \label{eigen(g)}
\widehat{V} = \sum_i \lambda_i |(e)i\rangle \langle(g)i| \;.
\end{equation}
Hence each non-zero value of $\lambda_i$ corresponds to a pair of states $%
|(g)i\rangle$ and $|(e)i\rangle$ that are coupled by $\widehat{V}$ and $%
\widehat{V}^{\dagger}$. In addition, the operator $\widehat{V}^{\dagger}%
\widehat{V}$ or $\widehat{V}\widehat{V}^{\dagger}$ may have eigenvalues
zero. The corresponding eigenstates are unaffected by the radiation field,
and they do not contribute to the coupling operator (\ref{eigen(g)}).

If the commutation rules (\ref{statementG}) are true, the steady-state
density matrices $\widehat{\rho }_{gg}$ and $\widehat{\rho }_{ee}$ are
diagonal on these bases. The diagonal elements $\pi _{i}^{(g)}$ and $\pi
_{j}^{(e)}$ are the stationary populations. Taking the diagonal elements of
the equations (\ref{pope}) and (\ref{popg}), we obtain the relations
\begin{equation}
\gamma \pi _{i}^{(e)}=-\gamma S\lambda _{i}^{2}\pi _{i}^{(e)}+\gamma
S\lambda _{i}^{2}\pi _{i}^{(g)}  \label{P(e)}
\end{equation}
\begin{equation}
-\gamma \sum_{j}{\cal W}_{ij}\pi _{j}^{(e)}=-\gamma S\lambda _{i}^{2}\pi
_{i}^{(g)}+\gamma S\lambda _{i}^{2}\pi _{i}^{(e)}\,  \label{P(g)}
\end{equation}
for the steady-state populations, with
\[
{\cal W}_{ij}=\sum_{q=0,\pm 1}\left| \langle (e)j|\widehat{D}%
_{q}|(g)i\rangle \right| ^{2}\,
\]
the probabilities of the spontaneous transitions $j\to i$. These transition
probabilities are normalized as $\sum_{i}{\cal W}_{ij}=1$ for all $j$. As is
seen from (\ref{P(e)}), if $\widehat{{\cal E}}_{e}$ has an eigenvalue equal
to zero, then the corresponding eigenstate $|(e)j\rangle $ is not populated.
Conversely, if one or more eigenvalues $\lambda _{i}^{2}$ of $\widehat{{\cal %
E}}_{g}$ are equal to zero, then a steady state exists where only the
corresponding eigenstates $|(g)i\rangle $ of the ground level are populated.
For $\lambda _{j}\neq 0$ it follows from (\ref{P(e)}) that the populations
of the ground- and excited-level substates are related by the equation
\begin{equation}
\pi _{j}^{(g)}=(1+\frac{1}{S\lambda _{j}^{2}})\pi _{j}^{(e)}\,.
\label{poplinks}
\end{equation}
From (\ref{P(g)}), it then follows that one can deduce a closed
system of equations for the excited-state populations in the form
\begin{equation}
\sum_{j}{\cal W}_{ij}\pi _{j}^{(e)}=\pi _{i}^{(e)}\,.  \label{spoint}
\end{equation}
These relations (\ref{spoint}) uniquely determine the populations $\pi
_{i}^{(e)}$, apart from normalization. This implies that the steady-state
density matrix of the excited level $\widehat{\rho }_{ee}$ depends on the
intensity and the detuning only through a normalization constant, which is a
function of the saturation parameter $S$. Moreover, they show that the
distribution over the excited-level substates can be considered as a
stationary point of the radiative relaxation operator, in the sense that
such a distribution is invariant under spontaneous decay to the ground level
\cite{taich98sp}.

\subsection{Condition for diagonal steady state}

\label{genform} The conjecture of the commutation relations (\ref{statementG}%
) can be formulated in an invariant form. It is sufficient to assume the
existence of two Hermitian operators $\widehat{E}$ and $\widehat{G}$, with $%
\widehat{E}$ acting on the excited states, and $\widehat{G}$ on the ground
states, and obeying the identities
\begin{equation}
\widehat{E}\widehat{V}=\widehat{V}\widehat{G}\;\;,\;\;\sum_{q=0,\pm 1}%
\widehat{D}_{q}^{\dagger }\widehat{E}\widehat{D}_{q}=\widehat{G}\;\;.
\label{EVG}
\end{equation}
From the first identity (\ref{EVG}) it follows that the operators $\widehat{E%
}$ and $\widehat{G}$ have an identical diagonal matrix form on the natural
bases. From this identity and its Hermitian conjugate $\widehat{V}^{\dagger }%
\widehat{E}=\widehat{G}\widehat{V}^{\dagger }$ one obtains the commutaton
rules
\begin{equation}
\lbrack \widehat{E},\widehat{V}\widehat{V}^{\dagger }]=0\;\;,\;\;[\widehat{G}%
,\widehat{V}^{\dagger }\widehat{V}]=0\;.  \label{commEVG}
\end{equation}
Starting from the relations (\ref{EVG}), while using the equations (\ref
{pope}) and (\ref{popg}), one easily derives that the steady-state
submatrices $\widehat{\rho }_{ee}$ and $\widehat{\rho }_{gg}$ are determined
by the relations
\begin{equation}
\widehat{\rho }_{ee}=\beta S\widehat{E}\;\;,\;\;\widehat{V}^{\dagger }%
\widehat{V}\widehat{\rho }_{gg}=\beta \left( 1+S\widehat{V}^{\dagger }%
\widehat{V}\right) \widehat{G}\;\;,  \label{sole}
\end{equation}
with $\beta $ a normalization constant. These relations are just the
operator expression of eq. (\ref{poplinks}). From (\ref{commEVG}) and (\ref
{sole}) it follows immediately that the commutation rules (\ref{statementG})
hold, and that the density matrix is diagonal on the natural basis.
Therefore, the problem of finding expressions for the steady-state density
matrix is now reduced to finding operators $\widehat{E}$ and $\widehat{G}$
that obey the relations (\ref{EVG}). These operators can be assumed to
depend only on the polarization vector ${\bf e}$, and not on the intensity
or the detuning. When operators $\widehat{E}$ and $\widehat{G}$ obeying (\ref
{EVG}) are found, the steady-state density matrix is determined by (\ref
{sole}), and it is indeed diagonal in the natural basis.

\section{Dark states}

As recalled in the Introduction, dipole transitions can be classified into
two classes depending on the occurrence of coherent population trapping
(CPT). For the first group, where CPT occurs, one or more of the eigenvalues
of the ground-state light-shift operator $\widehat{{\cal E}}_{g}$ vanish, so
that this operator cannot be inverted. During the optical-pumping process
atoms are accumulated in the corresponding eigenstates, which are termed
dark states, since they do not interact with the light field. Then the
trivial solution $\widehat{E}=\widehat{G}=0$ of the system (\ref{EVG}) still
determines a normalizable steady-state solution of the relations (\ref{sole}%
) obeying $\widehat{\rho }_{ee}=0$ and $\widehat{V}^{\dagger }\widehat{V}%
\widehat{\rho }_{gg}=0$. The ground-state density matrix $\widehat{\rho }%
_{gg}$ is composed of the ground-level dark states $|\Psi ^{(NC)}\rangle $,
which obey the equation
\begin{equation}
\widehat{V}\,|\Psi ^{(NC)}\rangle =0\,.  \label{cptcond}
\end{equation}
In order to specify the dark states in an invariant manner, we view state
vectors as tensors. A state vector $|\Psi \rangle $ in the ground level is
considered as a tensor $\Psi _{J_{g}}$ of the rank $J_{g}$, with (covariant)
components $\Psi _{J_{g}\,-\mu _{g}}$ specified by the expansion
\[
|\Psi \rangle =\sum_{\mu _{g}}(-1)^{-\mu _{g}}\Psi _{J_{g}\,-\mu _{g}}\left|
J_{g},\mu _{g}\right\rangle \;.
\]
Using the Wigner-Eckart theorem, we express the matrix elements of the
left-hand side of eq. (\ref{cptcond}) as
\begin{eqnarray*}
\langle J_{e},\mu _{e}|(\widehat{{\bf d}}\cdot {\bf e})|\Psi ^{(NC)}\rangle
&=&\langle J_{e}||d||J_{g}\rangle \sum_{q,\mu _{g}}C_{J_{g}\mu
_{g}\,1q}^{J_{e}\mu _{e}}(-1)^{-q}e_{-q}(-1)^{-\mu _{g}}\Psi _{J_{g}\,-\mu
_{g}}^{(NC)} \\
&=&\langle J_{e}||d||J_{g}\rangle (-1)^{-\mu _{e}}\{{\bf e}\otimes \Psi
_{J_{g}}^{(NC)}\}_{J_{e}\,-\mu _{e}}\;,
\end{eqnarray*}
where $\{\ldots \otimes \ldots \}$ denotes the standard definition of an
irreducible tensor product \cite{varsh75}, $\langle J_{e}||d||J_{g}\rangle $
is the reduced matrix element of the dipole moment operator, and as before, $%
e_{q}={\bf e}\cdot {\bf e}_{q}$ is the covariant spherical component of the
polarization vector ${\bf e}$. The invariant expression of eq. (\ref{cptcond}%
) in terms of a tensor product of rank $J_{e}$ reads
\begin{equation}
\{{\bf e}\otimes \Psi _{J_{g}}^{(NC)}\}_{J_{e}}=0\,,  \label{cptinv}
\end{equation}
with the normalization condition $({\Psi _{J_{g}}^{(NC)}}^{*}\cdot \Psi
_{J_{g}}^{(NC)})=1\,.$

\subsection{Transitions $J_g=J \rightarrow J_e=J$ with integer $J$}

For transitions with integer values of $J_g = J_e = J$ there is a single
dark state for any polarization \cite{smirnov89}. In order to write an
explicit and invariant form of $\Psi^{(NC)}_J$ in this case, we introduce
the $L$-fold tensor product of the vector ${\bf e}$ \cite
{manakov96,manakov97}
\begin{equation}  \label{aL}
\{{\bf e}\}_L=\{\ldots\{\{{\bf e}\otimes{\bf e}\}_2\otimes {\bf e}%
\}_3\,...\otimes{\bf e}\}_L\,.
\end{equation}
which are proportional to the spherical harmonics of a complex direction $%
n_{L M}({\bf e})$ (Appendix \ref{spher_harm}). Notice that the three
components $\{e\}_{1q}$ in the case $L=1$ coincide with the spherical
components of the (possibly complex) vector ${\bf e}$.

The Clebsch-Gordan expansion (\ref{cgexpansion}) for a product of two
spherical harmonics with the same argument leads to the result
\begin{equation}
\{\{{\bf e}\}_{L}\otimes \{{\bf e}\}_{J}\}_{K}=C_{L0\,J0}^{K0}\,\sqrt{\frac{%
L!\,J!\,(2K-1)!!\,({\bf e}\cdot {\bf e})^{L+J-K}}{K!\,(2L-1)!!\,(2J-1)!!}}\,\{%
{\bf e}\}_{K}\,.  \label{C-G}
\end{equation}
It follows from the symmetry of the Clebsch-Gordan coefficients that $%
C_{L0\,J0}^{K0}=0$ if $L+J-K$ is odd. Then choosing $L=1$ and $K=J$ we
obtain $\{{\bf e}\otimes \{{\bf e}\}_{J}\}_{J}=0$, so that the single dark
state as defined by (\ref{cptinv}) can be specified in tensor form as
\begin{equation}
\Psi _{J}^{(NC)}={\cal N}\{{\bf e}\}_{J}\,.  \label{NCJJ}
\end{equation}
The normalization constant follows from the equality
\begin{equation}
(\{{\bf e}\}_{J}\cdot \{{\bf e}^{*}\}_{J})=\frac{J!}{(2J-1)!!}({\bf e}\cdot
{\bf e})^{J/2}({\bf e}^{*}\cdot {\bf e}^{*})^{J/2}\,P_{J}\left( \frac{({\bf e%
}\cdot {\bf e}^{*})}{({\bf e}\cdot {\bf e})}\right) \,,  \label{eJeJ}
\end{equation}
as an example of the sum rule for spherical harmonics (\ref{sumrule}). Here $%
P_{J}(x)$ is the standard notation for Legendre polynomials. This leads to
the expression
\begin{equation}
{\cal N}=\left[ \frac{J!}{(2J-1)!!}\,({\bf e}\cdot {\bf e})^{J}\,P_{J}\left(
\frac{1}{({\bf e}\cdot {\bf e})}\right) \right] ^{-1/2}\;.  \label{A}
\end{equation}
In general, the algebraic and transformational properties of the dark state $%
\Psi _{J}^{(NC)}$ are the same as for spherical harmonics. The steady-state
density matrix
\begin{equation}
\widehat{\rho }_{gg}=|\Psi ^{(NC)}\rangle \langle \Psi ^{(NC)}|
\label{rhoJJ}
\end{equation}
obviously commutes with the light-shift operator $\widehat{{\cal E}}_{g}$.

In the special case of $J_{g}=J_{e}=1$, which is the prototype case of CPT
\cite{Aspect,invvscpt}, the dark state is specified by
\begin{equation}
|\Psi ^{(NC)}\rangle =\sum_{q=0,\pm 1}(-1)^{q}e_{-q}|1,\mu _{g}=q\rangle \;.
\label{dark1}
\end{equation}
It is well-known that the states $|1,q\rangle $ with angular momentum $1$
have the same transformation properties as the three spherical unit vectors $%
{\bf e}_{q}$, so that any state vector can be represented by a Cartesian
vector. When the state coupled to an excited state with angular momentum $1$
by the operator $\widehat{V}$, the vector representing the excited state is
represented by the vector that is the cross product of the ground-state
vector and the polarization vector, since the cross product is the only way
in which a vector can be formed from two vectors. Now a comparison of (\ref
{dark1}) with (\ref{polarization}) shows that the expansion coefficients are
identical, so that the state (\ref{dark1}) has the polarization vector ${\bf %
e}$ as its vector representation. This immediately explains why eq. (\ref
{cptcond}) holds for this state (\ref{dark1}), since the cross product of a
vector with itself vanishes \cite{invvscpt}. The explicit invariant form (%
\ref{NCJJ}) of the dark states for integer values of $J_{g}=J=J_{e}$
generalizes the well-known result for $J=1$.

\subsection{Transitions $J_g=J \rightarrow J_e=J-1$}

For transitions with $J_{e}=J_{g}-1$, the CPT-condition (\ref{cptinv}) takes
the form
\begin{equation}
\{{\bf e}\otimes \Psi _{J}^{(NC)}\}_{J-1}=0\,.  \label{jj-1}
\end{equation}
In this case there is a two-dimensional dark subspace, spanned by two
independent dark states \cite{smirnov89}. The scheme of light-induced
transitions in the natural coordinate frame is shown in Fig.2b. It is
explicitly seen that the dark state coincides with the outermost Zeeman
substate $|J,\mu =J\rangle $. The other linearly independent dark state is
determined analogously in the frame, connected with the second cylinder.

First we consider the case that $J$ is integer. An invariant tensorial
expression for the dark state is directly obtained when we notice that the
outermost Zeeman substate is given by the tensor $\left\{{\bf C}\right\}_J$
with ${\bf C}$ the circular component of the polarization vector in the
corresponding natural coordinate frame. This vector ${\bf C}$ is completely
specified by the requirements that it is normal to the polarization vector.
Hence, the solution of (\ref{jj-1}) is represented by the tensor
\begin{equation}  \label{psijj-1}
\Psi_J^{(NC)} = \left\{{\bf C}\right\}_J\; ,
\end{equation}
where
\begin{equation}  \label{invcond}
\left({\bf C}^*\cdot {\bf C}\right)=1 \;\;;\;\; \left({\bf C} \cdot {\bf C}
\right) = 0 \;\; ; \;\; \left({\bf e} \cdot {\bf C}\right) = 0 \; .
\end{equation}
The two independent (but not orthogonal) solutions are
\begin{equation}  \label{c12}
{\bf C}^{(1,2)} = \frac{\left[{\bf e}\times [{\bf e}\times{\bf e}%
^*]\right]\pm i[{\bf e}\times{\bf e}^*]\sqrt{({\bf e}\cdot {\bf e})}} {\sqrt{%
(1-|{\bf e}\cdot {\bf e}|^2)(1+|{\bf e}\cdot{\bf e}|)}} \, ,
\end{equation}
and the two corresponding dark states are called $\Psi_J^{(1)}$ and $%
\Psi_J^{(2)}$. These states are normalized and linearly independent but not
orthogonal. We can combine them into two orthogonal states defined by
\begin{equation}  \label{darkbasis}
\Psi^{(\pm)} = \frac{\Psi_J^{(1)}\pm\Psi_J^{(2)}} {\sqrt{2\left(1\pm({%
\Psi_J^{(1)}}^*\cdot\Psi_J^{(2)})\right)}}
\end{equation}
In order to calculate the dot product $\left( {\Psi_J^{(1)}}^*\cdot
\Psi_J^{(2)} \right)$ we take into account the relationship
\[
\left( \{{\bf a\}_J\cdot \{b\}_J }\right) = \frac{J!}{(2J-1)!!} \left(\sqrt{(%
{\bf a}\cdot{\bf a}) ({\bf b}\cdot{\bf b})}\right)^J\, P_J\left(\frac {({\bf %
a}\cdot{\bf b})} {\sqrt{({\bf a}\cdot{\bf a}) ({\bf b}\cdot{\bf b})}}%
\right)= ({\bf a}\cdot{\bf b})^J\,,
\]
which holds if either ${\bf a}$ or ${\bf b}$ is a circular vector. Hence we
obtain
\begin{equation}  \label{overlap}
\left( {\Psi_J^{(1)}}^*\cdot \Psi_J^{(2)} \right) = \left({{\bf C}^{(1)}}%
^*\cdot {\bf C}^{(2)}\right)^J= \left( \frac{1-|{\bf e}\cdot {\bf e}|}{1+|%
{\bf e}\cdot {\bf e}|}\right)^J
\end{equation}

Next we turn to the case of half-integer values of $J$. Then we can use the
correspondence between circular vectors and spinors. The tensor product of a
spinor $\chi$ (defined as a tensor of rank $1/2$) with itself into a tensor
of rank $1$ is always a circular vector, so that
\begin{equation}  \label{ctochi}
\left\{ \chi \otimes \chi \right\}_1 = {\bf C} \; ,
\end{equation}
with ${\bf C}\cdot{\bf C}=0$. The plane of this circular vector is normal to
the direction of the orientation of the spin vector represented by the
spinor. Conversely, any circular vector can be represented in the form (\ref
{ctochi}) for some spinor $\chi$. Now for a given polarization vector ${\bf e%
}$, the two circular vectors (\ref{c12}) correspond to two spinors $%
\chi^{(1,2)}$ so that
\begin{equation}  \label{ctochim}
\left\{ \chi^{(m)} \otimes \chi^{(m)} \right\}_1 = {\bf C}^{(m)} \;\;\; ,
\;\;\; m=1,2 \;\;.
\end{equation}
Since the two dark states $\Psi_J^{(m)}$ ($m=1,2$) are the outermost Zeeman
states in the two natural coordinate frames, they can be expressed in the
form
\begin{equation}  \label{x}
\Psi_J^{(m)} = \{ \chi^{(m)} \}_J \;\; ,
\end{equation}
where the tensor $\{\chi\}_J$ is constructed from $2J$ spinors $\chi$
\[
\{ \chi \}_J = \left\{ \ldots \left\{ \left\{ \chi \otimes \chi \right\}_1
\otimes \chi \right\}_{3/2} \ldots \otimes \chi \right\}_J \; .
\]
The orthonormalization is specified by equations (\ref{darkbasis}) and (\ref
{overlap}) for both integer and half-integer momenta.

In the steady state, the excited submatrix $\widehat{\rho }_{ee}$
disappears, and the ground-state density matrix $\widehat{\rho }_{gg}$ can
be an arbitrary density matrix within the two-dimensional subspace spanned
by the two dark states
\[
|\Psi _{\pm }^{(NC)}\rangle =\sum_{\mu _{g}}(-1)^{-\mu _{g}}\,\Psi
_{J_{g}\,-\mu _{g}}^{(\pm )}\,|J_{g},\mu _{g}\rangle \;.
\]
Obviously, any density matrix within this subspace commutes with the
light-shift operator $\widehat{{\cal E}}_{g}$. For any value of $J$, this
dark subspace depends only on the polarization vector ${\bf e}$. However,
the specific steady-state density matrix in which an atom will end up can
depend upon the initial state as well as on the intensity and the detuning
of the light field. This case of a transition with $J_{e}=J_{g}-1$ is the
only case of a dipole-allowed transition in which the steady state is not
unique.

\section{No dark states}

\subsection{General form of steady state}

For a transition without dark states, the steady-state solution is unique,
as has been proved in Ref. \cite{taich96j}. Then the excited level is
populated in the steady state. This is the case when the ground-state
light-shift operator $\widehat{{\cal E}}_{g}$ has no eigenvalues zero, so
that the operator $\widehat{V}^{\dagger }\widehat{V}$ acting within the $%
2J_{g}+1$ states of the ground level can be inverted. When operators $%
\widehat{E}$ and $\widehat{G}$ exist that obey the relations (\ref{EVG}),
the steady-state is obtained from (\ref{sole}) in the form
\begin{equation}
\widehat{\rho }_{ee}=\beta S\widehat{E}\;\;,\;\;\widehat{\rho }_{gg}=\beta
\left[ (\widehat{V}^{\dagger }\widehat{V})^{-1}+S\right] \widehat{G}
\label{rhonocpt}
\end{equation}
From the commutation rules (\ref{commEVG}) it follows that the
density matrix is diagonal on the natural basis. The optical
coherences are directly evaluated from eq. (\ref{current}), with the
result
\begin{equation}
\widehat{\overline{\rho }}_{eg}=\left( \widehat{\overline{\rho }}%
_{ge}\right) ^{\dagger }=\frac{\beta \Omega }{(\delta +i\gamma /2)}\widehat{V%
}(\widehat{V}^{\dagger }\widehat{V})^{-1}\widehat{G}\,.  \label{coh2}
\end{equation}
The constant $\beta $ follows from the normalization condition (\ref{trace}%
), and we obtain
\begin{equation}
\beta =\frac{1}{\alpha _{0}+2S\alpha _{1}}\,,  \label{beta}
\end{equation}
with the invariant expressions for the coefficients
\begin{equation}
\alpha _{1}={\rm Tr}\{\widehat{E}\}={\rm Tr}\{\widehat{G}\}  \label{alpha1}
\end{equation}
and
\begin{equation}
\alpha _{0}={\rm Tr}\left\{ (\widehat{V}^{\dagger }\widehat{V})^{-1}\widehat{%
G}\right\} \,.  \label{alpha0}
\end{equation}
These coefficients $\alpha _{0}$ and $\alpha _{1}$ depend on the
polarization only, not on the intensity or the detuning. The steady-state
density matrix depends in the intensity and the detuning only through the
value of the saturation parameter $S$, defined in (\ref{sat}).

It is noteworthy that the submatrix $\widehat{\rho}_{ee}$ of the excited
level is always proportional to the single operator $\widehat{E}$. This
implies that the steady-state anisotropy of the excited level, such as its
orientation and its alignment, is fully determined by the polarization
alone, independent of the saturation. The ground-level submatrix $\widehat{%
\rho}_{gg}$ is a linear combination of the two matrices $\widehat{G}$ and $(%
\widehat{V}^{\dagger}\widehat{V})^{-1}\widehat{G}$. For small values of the
saturation parameter, the steady-state submatrices are
\begin{equation}  \label{nosat}
\widehat{\rho}_{ee} = S \widehat{E}/\alpha_0\;\;,\;\; \widehat{\rho}_{gg} = (%
\widehat{V}^{\dagger}\widehat{V})^{-1}\widehat{G}/\alpha_0\;,
\end{equation}
whereas in the limit of strong saturation $S \rightarrow \infty$ we obtain
\begin{equation}  \label{satlim}
\widehat{\rho}_{ee} = \widehat{E}/2\alpha_1\;\;,\;\; \widehat{\rho}_{gg} =
\widehat{G}/2\alpha_1\;.
\end{equation}
The matrix for the optical coherence (\ref{coh2}) depends on the intensity
only through an overall factor $\beta \Omega$, that is equal to $\Omega /
\alpha_0$ in the low-intensity limit, and that approaches zero for strong
saturation.

\subsection{Transitions $J_g=J\rightarrow J_e=J+1$}

In order to find operators $\widehat{E}$ and $\widehat{G}$ that obey the
relations (\ref{EVG}) for transitions $J_g=J\rightarrow J_e=J+1$, we
introduce the operators
\begin{equation}  \label{vlmatr}
\widehat{V}^{ab}_{L}({\bf a}) = \sum_{M=-L}^{L} (-1)^M\widehat{T}^{ab}_{LM}
n_{L\;-M}({\bf a})\;,
\end{equation}
which are proportional to the dot product of the spherical harmonic $n_{LM}(%
{\bf a})$, introduced in (\ref{Yfunc}), and the tensor Wigner operator
\begin{equation}  \label{Twigner}
\widehat{T}_{LM}^{ab}=\sum_{\mu_a,\mu_b} |J_a,\mu_a\rangle (-1)^{J_b-\mu_b}
C^{LM}_{J_a\mu_a\;J_b\,-\mu_b} \langle J_b,\mu_b| \,.
\end{equation}
The indices $a$ and $b$ indicate the levels ($e$ or $g$), and ${\bf a}$ is a
vector in the complex three-dimensional space. We shall use the two
multiplication rules
\begin{equation}  \label{comm1}
\widehat{V}^{eg}_{1}({\bf e}) \widehat{V}^{ge}_{2J+1}({\bf e}) = \widehat{V}%
^{eg}_{2J+1}({\bf e}) \widehat{V}_{1}^{ge}({\bf e}) \;,\; \widehat{V}_1^{ge}(%
{\bf a})\widehat{V}_{2J+1}^{eg}({\bf b}) = \widehat{V}_{2J+1}^{ge}({\bf b})%
\widehat{V}_1^{eg}({\bf a}) \;.
\end{equation}
The first equation (\ref{comm1}) holds for any vector ${\bf e}$, and for
arbitrary value of the angular momenta $J_e$ and $J_g$. The second equation (%
\ref{comm1}) is valid for arbitrary complex vectors ${\bf a}$ and ${\bf b}$,
provided that $J_e = J_g + 1$. These relations follow from the
multiplication properties of the operators (\ref{vlmatr}) as given in
Appendix \ref{alg_vl}, in particular eqs. (\ref{first}) and (\ref{second2}).
In the notation of eq. (\ref{vlmatr}), the coupling operator is given by $%
\widehat{V}=\sqrt{({\bf e}\cdot{\bf e})(2J_e+1)/3}\, \widehat{V}^{eg}_{1}(%
{\bf e})$.

In order to identify the operators $\widehat{E}$ and $\widehat{G}$ we
introduce the short-hand notations
\begin{equation}  \label{Wmatr}
\widehat{V}^{eg}_{2J+1}({\bf e})=\widehat{W}\,;\;\;\; \widehat{V}%
^{ge}_{2J+1}({\bf e})=\widehat{\widetilde{W}}\,,
\end{equation}
In addition to $\widehat{V}$, we introduce the coupling operator
\begin{equation}  \label{Vmatr}
\widehat{\widetilde{V}} = \sqrt{({\bf e}\cdot{\bf e})(2J_e+1)/3} \, \widehat{%
V}^{ge}_{1}({\bf e})
\end{equation}
Notice that the operators $\widehat{W}$ and $\widehat{V}$ are raising
operators, which map substates of the ground level onto excited states. The
operators $\widehat{\widetilde{W}}$ and $\widehat{\widetilde{V}}$ are
lowering operators.

Using the relations (\ref{comm1}) one can find operators $\widehat{E}$ and $%
\widehat{G}$ with the properties (\ref{EVG}). They are specified by the
definitions
\begin{equation}
\widehat{E}=\widehat{W}\widehat{W}^{\dagger }\;\;,\;\;\widehat{G}=\widehat{%
\widetilde{W}}\widehat{\widetilde{W}}^{\dagger }\,.  \label{EG}
\end{equation}
With the notation (\ref{Wmatr}), and the substitution ${\bf a}={\bf e}$, $%
{\bf b}={\bf e}^{*}$ in (\ref{comm1}), we obtain the identities
\begin{equation}
\widehat{V}\widehat{\widetilde{W}}=\widehat{W}\widehat{\widetilde{V}}\;,\;%
\widehat{\widetilde{V}}\widehat{\widetilde{W}}^{\dagger }=\widehat{W}%
^{\dagger }\widehat{V}\;,  \label{VW}
\end{equation}
which indeed prove the first equation (\ref{EVG}):
\begin{equation}
\widehat{E}\widehat{V}=\widehat{W}\widehat{W}^{\dagger }\widehat{V}=\widehat{%
W}\widehat{\widetilde{V}}\widehat{\widetilde{W}}^{\dagger }=\widehat{V}%
\widehat{\widetilde{W}}\widehat{\widetilde{W}}^{\dagger }=\widehat{V}%
\widehat{G}\,.  \label{proof1}
\end{equation}
The second equation (\ref{EVG}) is easily verified when the summation is
performed over three Cartesian polarization vectors ${\bf e}_{i}$. By using
the second identity (\ref{comm1}), with ${\bf e}_{i}$ substituted for ${\bf e%
}$, and ${\bf e}$ for ${\bf b}$, one finds
\begin{equation}
\sum_{q=0,\pm 1}\widehat{D}_{q}^{\dagger }\widehat{E}\widehat{D}%
_{q}=\sum_{q=0,\pm 1}\widehat{D}_{q}^{\dagger }\widehat{W}\widehat{W}%
^{\dagger }\widehat{D}_{q}=\widehat{\widetilde{W}}\sum_{q=0,\pm 1}\widehat{D}%
_{q}\widehat{D}_{q}^{\dagger }\;\widehat{\widetilde{W}}^{\dagger }=\widehat{G%
}  \label{proof2}
\end{equation}
where eq. (\ref{norm}) is used in the last step. As indicated in Sec. \ref
{genform}, this also proves the commutation rules (\ref{statementG}) for the
steady state in the present class of transitions.

Now that we have identified the operators $\widehat{E}$ and $\widehat{G}$
with the desired properties, the steady-state density matrix is directly
obtained in the form (\ref{rhonocpt}) and (\ref{coh2}). We use the first
equation (\ref{EVG}), combined with the commutation relation (\ref{commEVG})
for $\widehat{G}$, and we introduce the operator
\begin{equation}  \label{Xdef}
\widehat{X} = (\widehat{V}^{\dagger}\widehat{V})^{-1} \widehat{V}^{\dagger}
\widehat{W}\;,
\end{equation}
acting on the states of the ground level. This leads to the expressions
\begin{eqnarray}  \label{rhoJJ+1}
\widehat{\rho}_{ee} &=& \beta S \widehat{W} \widehat{W}^{\dagger}  \nonumber
\\
\widehat{\rho}_{gg} &=& \beta \left(\widehat{X}\widehat{X}^{\dagger} +S%
\widehat{\widetilde{W}}\widehat{\widetilde{W}}^{\dagger}\right) \\
\widehat{\overline{\rho}}_{eg} &=& \left(\widehat{\overline{\rho}}%
_{ge}\right)^{\dagger} = \frac{\beta\Omega}{\delta+i\gamma/2} \widehat{W}%
\widehat{X}^{\dagger} \; ,  \nonumber
\end{eqnarray}
Alternatively, the operator $\widehat{X}$ is defined by the relation $%
\widehat{V}\widehat{X}=\widehat{W}$. The density matrix is indeed diagonal
on the natural basis, and the commutation rules (\ref{statementG}) hold.

It is illuminating to express the various operators occurring in the
equations (\ref{rhoJJ+1}) in terms of basisvectors. The commutation rules (%
\ref{commEVG}) can be expressed as
\begin{equation}
\lbrack \widehat{W}\widehat{W}^{\dagger },\widehat{V}\widehat{V}^{\dagger
}]=0\;\;,\;\;[\widehat{\widetilde{W}}\widehat{\widetilde{W}}^{\dagger },%
\widehat{V}^{\dagger }\widehat{V}]=0\;.  \label{commVW}
\end{equation}
The operators $\widehat{V}$ and $\widehat{W}$ depend on the polarization
vector ${\bf e}$. From the definition (\ref{vlmatr}) one obtains the
relation
\begin{equation}
\lbrack \widehat{V}_{L}^{ab}({\bf a})]^{\dagger }=(-1)^{J_{a}-J_{b}}\widehat{%
V}_{L}^{ba}({\bf a}^{*})\;,  \label{herm}
\end{equation}
which shows that, apart from a minus sign, the Hermitian conjugates of $%
\widehat{\widetilde{W}}$ and $\widehat{\widetilde{V}}$ are equal to the
operators $\widehat{W}$ and $\widehat{V}$ with the polarization vector
replaced by its complex conjugate ${\bf e}^{*}$. Hence, when the
polarization vector is taken as ${\bf e}^{*}$ rather than ${\bf e}$, the
expressions (\ref{rhoJJ+1}) and (\ref{Xdef}) hold with the replacements $%
\widehat{V}\leftrightarrow \widehat{\widetilde{V}}^{\dagger }$ and $\widehat{%
W}\leftrightarrow \widehat{\widetilde{W}}^{\dagger }$. The commutation rules
analogous to (\ref{commVW}) are then
\begin{equation}
\lbrack \widehat{\widetilde{W}}^{\dagger }\widehat{\widetilde{W}},\widehat{%
\widetilde{V}}^{\dagger }\widehat{\widetilde{V}}]=0\;\;,\;\;[\widehat{W}%
^{\dagger }\widehat{W},\widehat{\widetilde{V}}\widehat{\widetilde{V}}%
^{\dagger }]=0\;.  \label{commtildeVW}
\end{equation}
The first commutation rule (\ref{commVW}) shows that the natural basis
states $|(e)j\rangle $, defined as eigenstates of $\widehat{V}\widehat{V}%
^{\dagger }$, are also eigenstates of $\widehat{W}\widehat{W}^{\dagger }$,
and the (positive) eigenvalues are called $\nu _{j}^{2}$, with $\nu _{j}$
positive. In complete analogy to the relation (\ref{diagV2}) between the
states $|(g)i\rangle $ and $|(e)j\rangle $ as coupled by $\widehat{V}$, we
notice that the states $\widehat{W}^{\dagger }|(e)i\rangle $ are eigenstates
of the ground-level operator $\widehat{W}^{\dagger }\widehat{W}$, with the
same eigenvalue $\nu _{i}^{2}$. Hence, we denote the normalized eigenstates
as $\widetilde{|(g)i\rangle }$, so that $\widehat{W}$ can be expanded as
\begin{equation}
\widehat{W}=\sum_{i}\nu _{i}|(e)i\rangle \widetilde{\langle (g)i|}\;.
\label{eigenW}
\end{equation}
From the second commutation rule in (\ref{commtildeVW}) it follows
that the ground-level states $\widetilde{|(g)i\rangle }$ are
eigenstates of the operator
$\widehat{\widetilde{V}}\widehat{\widetilde{V}}^{\dagger }$, which
means that they form the natural basis for the ground level for the
polarization ${\bf e}^{*}$. From the second identity (\ref{commVW})
one finds by the same argument that the states
$\widehat{\widetilde{W}}^{\dagger
}|(g)i\rangle $ are eigenstates of $\widehat{\widetilde{W}}^{\dagger }%
\widehat{\widetilde{W}}$. From the first identity (\ref{commtildeVW}) it
follows that these states form the natural excited basis for the
polarization ${\bf e}^{*}$, which we indicate as $\widetilde{|(e)i\rangle }$%
. Hence, the operators $\widehat{\widetilde{V}}$ and $\widehat{\widetilde{W}}
$ can be explicitly expressed as
\begin{equation}
\widehat{\widetilde{V}}=\sum_{i}\lambda _{i}\widetilde{|(g)i\rangle }%
\widetilde{\langle (e)i|}\;,\;\widehat{\widetilde{W}}=\sum_{i}\nu
_{i}|(g)i\rangle \widetilde{\langle (e)i|}\;.  \label{eigentildeVW}
\end{equation}
For symmetry reasons, the values $\lambda _{i}$ and $\nu _{i}$ must be the
same as in eqs. (\ref{eigen(g)}) and (\ref{eigenW}). The operator $\widehat{X%
}$ defined in (\ref{Xdef}) can be expanded as
\begin{equation}
\widehat{X}=\sum_{i}(\nu _{i}/\lambda _{i})|(g)i\rangle \widetilde{\langle
(g)i|}\;.  \label{eigenX}
\end{equation}
The steady-state populations of the natural basis states follow from the
expressions (\ref{rhoJJ+1}), with the result
\begin{equation}
\pi _{i}^{(e)}=\beta \nu _{i}^{2}S\;,\;\pi _{i}^{(g)}=\beta \nu
_{i}^{2}(\lambda _{i}^{-2}+S)\;.
\end{equation}
The relations (\ref{poplinks}) determine the ratio between the population of
an excited state of the natural basis set and the corresponding ground
state. The populations of different excited states of the natural basis set
are proportional to the eigenvalues $\nu _{i}^{2}$.

In the special case of linear polarization, the polarization vectors ${\bf e}
$ and ${\bf e}^*$ are identical. When we take the quantization axis in the
polarization direction, the natural basis of states coincide with the Zeeman
states $|J_g, \mu \rangle$ which is only coupled to the excited state $|J_e,
\mu \rangle$. The operator $\widehat{W}$ is proportional to the spherical
temsor $\widehat{T}^{eg}_{2J+1\,0}$, and the corresponding eigenvalues are
proportional to the Clebsch-Gordan coefficients $\nu_{\mu} \propto
C^{2J+1\,0}_{J+1\,\mu\;J\,-\mu}$. The resulting excited-state populations $%
\propto |C^{2J+1\,0}_{J+1\,\mu\;J\,-\mu}|^2$ for this special case of linear
polarization driving a transition $J \rightarrow J+1$ have been indicated
before in ref. \cite{macek74}.

Calculations of matrix elements of the operators $\widehat{W}$, $\widehat{%
\widetilde{W}}$ and $\widehat{X}$ are given in the appendix (\ref{matr_el})
for the natural coordinate frame. However, more physical insight can be
obtained by expanding the operators $\widehat{X}$ in the form of an
invariant superposition of the operators $\widehat{V}^{gg}_{L}$
\begin{equation}  \label{Xtovl}
\widehat{X}=\sqrt{\frac 3{({\bf e}\cdot{\bf e})(2J+3)}} \sum_{L=0}^{2J} C_L
\widehat{V}^{gg}_{L}({\bf e})\,.
\end{equation}
In order to find the coefficients $C_L$ we use the property (\ref{vl_cge}):
\begin{eqnarray}  \label{comp1L_JJ+1}
&&\widehat{V}^{eg}_{1}({\bf e})\widehat{V}^{gg}_{L}({\bf e}) =
\sum_{K=L-1}^{L+1} E(L,K) \widehat{V}^{eg}_{K}({\bf e})  \nonumber \\
&&E(L,K)=(-1)^{2J+L}\sqrt{3(2L+1)} \left\{
\begin{array}{rcl}
K & 1 & \;\;\; L \\
J & J & J+1
\end{array}
\right\} C^{K0}_{1 0\;L 0} \,.
\end{eqnarray}
The recurrent equations for the coefficients $C_L$ follows from the defining
relation $\widehat{V}\widehat{X}=\widehat{W}$
\begin{equation}  \label{recCJJ+1}
E(L-1,L)C_{L-1}+E(L+1,L)C_{L+1}=\delta_{L,2J+1}\,, \;\;\; L=0,1, \ldots,
2J+1 \,.
\end{equation}
Depending on whether $J$ is an integer or a halfinteger, the odd or even
coefficients are equal to zero. In both cases the nonzero coefficients $C_L$
are written as
\begin{equation}  \label{solCJJ+1}
C_L = \sqrt{\frac{(2L+1)(2J+3)}{3(2J+1)}\frac{(2J-L)!(2J+L+1)!}{(4J+1)!}} \;.
\end{equation}
A remarkable peculiarity of the solution (\ref{solCJJ+1}), which becomes
manifest at large angular momentum $J$, is a rapid decrease of the
coefficients $C_L$ with a decrease of index $L$. This fact allows
approximate calculations by restricting the expansion (\ref{Xtovl}) to only
a few terms. For example, the ratio
\[
\frac{C_{2J}}{C_{2J-2}}=(4J+1)\sqrt{\frac{2J}{4J-3}} \approx 2\sqrt{2}\,J
\]
for $J=4$ (that corresponds to the cycling transition of the $D_2$-line of $%
^{133}$Cs) amounts to about $13.3$. As a result, if we use the approximation
\[
\widehat{X} \approx \sqrt{\frac 3{({\bf e}\cdot{\bf e})(2J+3)}}\, C_{2J}
\widehat{V}^{gg}_{2J}({\bf e})\,,
\]
then in calculating quantities as the population or the orientation of the
levels and the average dipole moment, the error remains below one percent.

The normalization constants $\alpha_i$ can be found in explicit form for
arbitrary $J$. From the sum rule for spherical harmonics (\ref{sumrule}) it
follows that
\begin{equation}  \label{alpha1JJ+1}
\alpha_1 = \mbox{Tr}\left\{\widehat{W}\widehat{W}^{\dagger}\right\} = P_{2
J+1}\left(\frac 1{({\bf e}\cdot{\bf e})}\right) \;.
\end{equation}
Using the expansion (\ref{Xtovl}), we find
\begin{equation}  \label{alpha0JJ+1}
\alpha_0 = \mbox{Tr}\left\{\widehat{X}\widehat{X}^{\dagger}\right\} = \frac{3%
}{({\bf e}\cdot{\bf e})(2J+3)} \sum_{L=0}^{2J} C_L^2 P_L\left(\frac 1{({\bf e%
}\cdot{\bf e})}\right)\,.
\end{equation}
The coefficients $\alpha_0$ and $\alpha_1$ are simultaneously even (for $J$
a halfinteger) or odd (for $J$ an integer). As one expects, the populations
of the levels determined by the ratio of $\alpha_0$ and $\alpha_1$ do not
depend on the sign of $({\bf e}\cdot{\bf e})$.

\subsection{Transitions $J_g=J\rightarrow J_e=J$ with $J$ a halfinteger}

For a transition between levels with equal half-integer values $J$ of the
angular momenta, the coupling operator $\widehat{V}$ on the basis of the
Zeeman substates is represented by a square matrix. This implies that the
concept of the inverse operator $(\widehat{V})^{-1}$ as corresponding to the
inverse matrix can be used, in the sense that $(\widehat{V})^{-1}\widehat{V}
= \widehat{\Pi}_g$, $\widehat{V}(\widehat{V})^{-1} = \widehat{\Pi}_e$. It is
nearly trivial to find operators $\widehat{E}$ and $\widehat{G}$ that obey
the conditions (\ref{EVG}) for this class of transitions. It is immediately
obvious that these conditions are satisfied with the choice $\widehat{E}=%
\widehat{\Pi}_e$ and $\widehat{G}=\widehat{\Pi}_g$, in view of the identity (%
\ref{isotrop}). Then expressions (\ref{rhonocpt}) and (\ref{coh2}) for the
density matrix take the form
\begin{eqnarray}  \label{rhoJJh}
\widehat{\rho}_{ee} &=& \beta S \widehat{\Pi}_e  \nonumber \\
\widehat{\rho}_{gg} &=& \beta\left[ (\widehat{V}^{\dagger}\widehat{V})^{-1}
+ S \widehat{\Pi}_g \right] = \beta\left[ (\widehat{V})^{-1}(\widehat{V}%
^{\dagger})^{-1} + S \widehat{\Pi}_g \right] \\
\widehat{\overline{\rho}}_{eg} &=& \left(\widehat{\overline{\rho}}%
_{ge}\right)^{\dagger} =\frac{\beta\Omega}{(\delta+i\gamma/2)}(\widehat{V}%
^{\dagger})^{-1}\,.  \nonumber
\end{eqnarray}
The commutation rules (\ref{commEVG}) are trivially obeyed, so that the
density matrix is diagonal on the natural bases. These expressions have been
obtained in Ref. \cite{taich95}. Remarkably, eq. (\ref{rhoJJh}) shows that
for this class of transitions, the excited-state density matrix $\widehat{%
\rho}_{ee}$ is always isotropic at arbitrary field parameters. The
ground-state density matrix $\widehat{\rho}_{gg}$ consists of two parts. the
term proportional to $(\widehat{V}^{\dagger}\widehat{V})^{-1}$ is
anisotropic, and describes the distribution among Zeeman substates in the
low-saturation limit $S\ll1$. The isotropic part is dominant in the limit of
strong saturation (\ref{satlim}).

In fact, the expressions (\ref{rhoJJh}) can also be represented in the form (%
\ref{rhoJJ+1}), in terms of operators $\widehat{W}$, $\widehat{\widetilde{W}}
$ and $\widehat{X}$. In the same spirit as eq. (\ref{Wmatr}), for the
present class of transitions we introduce the operators
\begin{equation}  \label{WmatrJJ}
\widehat{V}^{eg}_{0}({\bf e})=\widehat{W}\,;\;\;\; \widehat{V}^{ge}_{0}({\bf %
e})=\widehat{\widetilde{W}}\,,
\end{equation}
where now the rank of the operators takes the minimal value $0$. With this
definition, we can maintain the expressions (\ref{EG}) for the operators $%
\widehat{E}$ and $\widehat{G}$, which are indeed isotropic. With the
definition (\ref{Vmatr}) of $\widehat{\widetilde{V}}$, the identities (\ref
{VW})-(\ref{proof2}) remain valid in the present case. This follows from
equations (\ref{first}) and (\ref{second1}) of Appendix \ref{alg_vl}. With
the definition (\ref{Xdef}) of the operator $\widehat{X}$, the proof of eqs.
(\ref{rhoJJ+1}) as well as of the expansions (\ref{eigenW})-(\ref{eigenX})
can be carried over directly. The main simplification is that in the present
case of a transition $J_g = J \rightarrow J_e = J$, the coefficients $\nu_i$
are all the same, while the natural basis states $|(e)i \rangle$ and $%
\widetilde{|(g)i\rangle}$ have an identical vector form. Therefore, on the
basis of Zeeman states of the excited and the ground level, the raising
operators $\widehat{W}$ and $\widehat{\widetilde{W}}$ have the form of a
unit matrix.

The matrix elements of the inverse operators $(\widehat{V})^{-1}$, $(%
\widehat{V}^{\dagger})^{-1}$ and $(\widehat{V}^{\dagger}\widehat{V})^{-1}$
can be found in a closed analytical form in the natural coordinate frame for
all $J$ (see Appendix \ref{matr_el}). However, for further applications, it
will be convenient to express $(\widehat{V})^{-1}$ in an invariant form in
terms of the spherical harmonics $n_{L M}({\bf e})$ defined in (\ref{Yfunc}%
). Since the coupling operator $\widehat{V}$ is proportional to the operator
(\ref{vlmatr}) with the rank $L=1$, we can write in the present case
\[
(\widehat{V})^{-1} =\sqrt{\frac 3{({\bf e}\cdot{\bf e})(2J+1)}}\; (\widehat{V%
}^{eg}_{1}({\bf e}))^{-1} \,.
\]
We use the equations
\begin{eqnarray}  \label{comp1L}
&&\widehat{V}^{eg}_{1}({\bf e})\widehat{V}^{ge}_{L}({\bf e}) =
\sum_{K=L-1}^{L+1} E(L,K) \widehat{V}^{ee}_{K}({\bf e}) \;,  \nonumber \\
&&E(L,K)=(-1)^{2J+L+1}\sqrt{3(2L+1)} \left\{
\begin{array}{rcl}
K & 1 & L \\
J & J & J
\end{array}
\right\} C^{K0}_{1 0\;L 0} \,,
\end{eqnarray}
which is a special case of (\ref{vl_cge}). The operator $(\widehat{V}%
^{eg}_{1}({\bf e}))^{-1}$ is expanded in the operators $\widehat{V}^{ge}_{L}(%
{\bf e})$ according to
\begin{equation}  \label{vl_expansion}
(\widehat{V}^{eg}_{1}({\bf e}))^{-1}= \sum_{L=0}^{2J} C_L \widehat{V}%
^{ge}_{L}({\bf e})\,.
\end{equation}
In order to find the coefficients $C_L$ we substitute this expansion into
the identity
\[
\widehat{V}^{eg}_{1}({\bf e})(\widehat{V}^{eg}_{1}({\bf e}))^{-1} = \widehat{%
\Pi}_e\;.
\]
While noting that $E(L,L)=0$ (since $C^{L0}_{1 0\;L 0}=0$) one arrives at
the two-term recurrent relation
\begin{equation}  \label{recCJJh}
E(L-1,L)C_{L-1}+E(L+1,L)C_{L+1}=\delta_{L,0}\sqrt{2J+1}\,, \;\;\; L=0,1,
\ldots, 2J \,.
\end{equation}
Expressions for the coefficients with odd indices follow from the relations (%
\ref{recCJJh}), with the result
\begin{equation}  \label{solCJJh}
C_L = (-1)^{\frac{L-1}{2}} \frac{(L-1)!!}{L!!}\sqrt{\frac{%
(2L+1)2J(2J+1)(2J+2)}{3} \frac{(2J+L)!!(2J-L-1)!!}{(2J-L)!!(2J+L+1)!!}} \;\;
\end{equation}
This is a sequence of terms with alternating signs, which slowly grows in
absolute value with the index $L$. It follows from (\ref{recCJJh}) that the
coefficients $C_L$ with even index are equal to zero.

The normalization coefficient (\ref{alpha1}) is obviously
\begin{equation}
\alpha _{1}={\rm Tr}\{\widehat{\Pi }_{e}\}=2J+1\,.  \label{alpha1JJ}
\end{equation}
The other normalization coefficient $\alpha _{0}={\rm Tr}\{(\widehat{V}%
^{\dagger }\widehat{V})^{-1}\}$ follows from the expansion (\ref
{vl_expansion}), the orthogonality of the Wigner operators and the sum rule
for spherical harmonics (\ref{sumrule}), with the result
\begin{equation}
\alpha _{0}=\frac{3}{({\bf e}\cdot {\bf e})(2J+1)}\sum_{L=1,3,\ldots
}^{2J}C_{L}^{2}P_{L}\left( \frac{1}{({\bf e}\cdot {\bf e})}\right) \,.
\label{alpha0JJ}
\end{equation}
This expression contains only even powers of $({\bf e}\cdot {\bf e})$.

\section{Excited-state population and AC Stark shift}

In this section, we study the polarization dependence of the total
excited-state population. This determines, for example, the total
fluorescence and thereby also the total absorption in an atomic vapor. For
unpolarized atoms and in the low-saturation limit, the total excited-state
population is given by $\pi _{e}^{(unpol)}=(S/3)(2J_{e}+1)/(2J_{g}+1)\,$,
independent of the polarization. This is the case of linear absorption. We
now consider as an application of the results of the previous section the
total excited-state population $\pi _{e}={\rm Tr}\{\widehat{\rho }_{ee}\}$
in the steady state. When we calculate the trace of (\ref{rhonocpt}), while
taking into account normalization (\ref{trace}), we find
\begin{equation}
\pi _{e}=\frac{S\,\alpha _{1}/\alpha _{0}}{1+2\,S\,\alpha _{1}/\alpha _{0}}%
\;,  \label{pie}
\end{equation}
where the real functions $\alpha _{1}(\varepsilon )$ and $\alpha
_{0}(\varepsilon )$ are expressed in (\ref{alpha1JJ+1}) and (\ref{alpha0JJ+1}%
) for the transition $J\rightarrow J+1$, and by (\ref{alpha1JJ}) and (\ref
{alpha0JJ}) for the transition $J\rightarrow J$ (recall that ${\bf e}\cdot
{\bf e}=\cos (2\varepsilon )$). As is seen from (\ref{pie}), the analytical
expression for the total excited-state population is similar as for a
two-level atom with non-degenerate levels (see e.g. \cite{eberly78}) but
with an effective saturation parameter $\widetilde{S}=S\,\alpha _{1}/\alpha
_{0}$, which depends on the class of transitions and on the light
polarization. If the saturation intensity $I_{sat}$ is defined as the
intensity at which $\pi _{e}=1/4$ (at zero detuning), then we obtain
\begin{equation}
I_{sat}(\varepsilon )=\frac{\alpha _{0}(\varepsilon )}{\alpha
_{1}(\varepsilon )}\,I_{0}\;;\;\;\;I_{0}=\frac{2\pi ^{2}}{3}\,\frac{\hbar
c\gamma }{\lambda ^{3}}\;,  \label{Isat}
\end{equation}
where $I_{0}$ is the usual measure for the saturation intensity. This shows
that now the excited-state population, and thereby the absorption, depends
on the polarization.

This is demonstrated in ~Fig.  \ref{fig3}a. for the class of transitions $%
J\to J$ with $J$ a half-integer. This shows that the total absorption
cross-section is reduced compared with the case of linear absorption, so
that the medium becomes more transparent due to optical pumping. The
reduction is the lowest for linear polarization, and it is complete for
circular polarization. Moreover, the reduction increases with the value of $J
$. Recall that for CPT transitions, $J\to J-1$ and $J\to J$ with $J$ an
integer, the stationary excited-state population is obviously zero, and the
transparency is complete. The decrease in absorption indicates that the
atoms are pumped to states that are more weakly coupled than average.
Roughly speaking, this implies that pairs of coupled states $|(e)i\rangle $
and $|(g)i\rangle $ for which the total steady-state population is large,
tend to have a relatively small coupling constant $\lambda _{i}^{2}$. Since
the form (\ref{Eg}) of the light shift operators shows that the states $%
|(e)i\rangle $ and $|(g)i\rangle $ are shifted by $-\delta S\lambda _{i}^{2}$
and $\delta S\lambda _{i}^{2}$, we may also conclude that this class of
transitions $J\to J$ with $J$ a half-integer tends to pump the atoms to
states with lower AC Stark shift.

In contrast, for $J\to J+1$ transitions the absorption is enhanced by
optical pumping. This is shown in Fig. ~\ref{fig3}b. Again, the effect of
optical pumping on the total absorption increases with $J$, and it increases
also when the polarization is varied from linear to circular. These results
are related to the recently discussed effect of electromagnetically induced
absorption (EIA) under two-frequency excitation in the Hanle configuration
on $J\to J+1$ transitions \cite{EIA}. Indeed, when the frequencies coincide
or the magnetic field is zero, we have a situation close to the stationary
interaction of atoms with elliptically polarized light considered here. When
the frequency difference or the Zeeman splitting is sufficiently large,
significant ground-state depolarization appears, and the absorption should
be close to the linear absorption of unpolarized atoms. In this sense, the
enhancement of absorption by optical pumping as illustrated in Fig.~\ref
{fig3}.b resembles EIA. By the same argument as used above, we conclude that
$J\to J+1$ transitions tend to pump the atoms to state with larger AC Stark
shifts. This tendency was found before in special cases in the context of
sub-Doppler laser cooling by polarization gradients \cite{Dalibard}.

\section{Broad-band radiation}

So far, we discussed the steady-state solutions of the generalized optical
Bloch equations (\ref{gobe1})-(\ref{gobe4}), which describe an atomic
transition driven by monochromatic polarized light. In this chapter we point
out that the results can be generalized to the case of light with a finite
bandwidth. Broad-band radiation is described by modeling the electric field
as a stationary stochastic process. The dynamics of an atom in such a field
with central frequency $\omega $ is described by the same eqs (\ref{gobe1})-(%
\ref{gobe4}), where now the Rabi frequency $\Omega (t)$ is a complex-valued
function of time, proportional to the positive-frequency part of the
fluctuating electric field. We are interested in the steady-state stochastic
average of the submatrices $\widehat{\rho }_{ee}$ and $\widehat{\rho }_{gg}$%
. The time-dependent solution of eq. (\ref{gobe1}) is
\begin{equation}
\widehat{\overline{\rho }}_{eg}(t)=\int_{0}^{\infty }d\tau \exp [(i\delta
-\gamma /2)\tau ]i\Omega (t-\tau )\left[ \widehat{V}({\bf e})\widehat{\rho }%
_{gg}(t-\tau )-\widehat{\rho }_{ee}(t-\tau )\widehat{V}({\bf e})\right] .
\label{stoch}
\end{equation}
When we substitute this expression and its analogue for $\widehat{\overline{%
\rho }}_{ge}(t)=\widehat{\overline{\rho }}_{eg}^{\dagger }(t)$ into eqs. (%
\ref{gobe3}) and (\ref{gobe4}), we arrive at a pair of stochastic
integro-differential equations for the submatrices $\widehat{\rho }_{ee}(t)$
and $\widehat{\rho }_{gg}(t)$. The r.h.s. of these equations contains the
stochastic parts \newline
$\Omega ^{*}(t)\Omega (t-\tau )\widehat{\rho }_{gg}(t-\tau ) $, $\Omega
^{*}(t)\Omega (t-\tau )\widehat{\rho }_{ee}(t-\tau )$ and their Hermitian
conjugates. Stochastic averaging of these equations leads to a closed set of
equations for the steady-state stochastic averages $\left\langle \widehat{%
\rho }_{ee}\right\rangle $ and $\left\langle \widehat{\rho }%
_{gg}\right\rangle $, provided that the stochastic average of these terms
may be factorized as
\begin{equation}
\left\langle \Omega ^{*}(t)\Omega (t-\tau )\widehat{\rho }_{gg}(t-\tau
)\right\rangle =\left\langle \Omega ^{*}(t)\Omega (t-\tau )\right\rangle
\left\langle \widehat{\rho }_{gg}\right\rangle ,  \label{stochav}
\end{equation}
and similarly for the other terms. When we substitute this factorized form
of the type (\ref{stochav}) in the equations for $\widehat{\rho }_{ee}$ and $%
\widehat{\rho }_{gg}$, we arrive in the steady state at closed equations for
the stochastic averages $\left\langle \widehat{\rho }_{ee}\right\rangle $
and $\left\langle \widehat{\rho }_{gg}\right\rangle $%
\begin{equation}
\gamma \left\langle \widehat{\rho }_{ee}\right\rangle =-R\{\widehat{V}%
\widehat{V}^{\dagger },\left\langle \widehat{\rho }_{ee}\right\rangle \}+2R%
\widehat{V}\left\langle \widehat{\rho }_{gg}\right\rangle \widehat{V}%
^{\dagger }+iL[\widehat{V}\widehat{V}^{\dagger },\left\langle \widehat{\rho }%
_{ee}\right\rangle ],  \label{popestoch}
\end{equation}
\begin{equation}
-\gamma \sum_{q=0,\pm 1}\widehat{D}_{q}^{\dagger }\left\langle \widehat{\rho
}_{ee}\right\rangle \widehat{D}_{q}=-R\{\widehat{V}^{\dagger }\widehat{V}%
,\left\langle \widehat{\rho }_{gg}\right\rangle \}+2R\widehat{V}^{\dagger
}\left\langle \widehat{\rho }_{ee}\right\rangle \widehat{V}-iL[\widehat{V}%
^{\dagger }\widehat{V},\left\langle \widehat{\rho }_{gg}\right\rangle ].
\label{popgstoch}
\end{equation}
The two real parameters $R$ and $L$ are defined by the equation
\begin{equation}
R+iL=\int_{0}^{\infty }d\tau \exp [(i\delta -\gamma /2)\tau ]\left\langle
\Omega ^{*}(t)\Omega (t-\tau )\right\rangle .  \label{stochsat}
\end{equation}
The quantity $R$ is a measure of the stimulated transition rates, and $L$
determines the strength of the light shift. When we substitute this
factorized form of the type (\ref{stochav}) in the equations for $\widehat{%
\rho }_{ee}$ and $\widehat{\rho }_{gg}$.

The factorization (\ref{stochav}) is exact in the special case that the
finite bandwidth is due to phase fluctuations only. In that case we can
write $\Omega (t)=\Omega \exp (-i\psi (t))$, with a stochastic phase $\psi $%
. The factorization is then justified since the phase change $\psi (t)-\psi
(t-\tau )$ in the time interval $[t-\tau ,t]$ can safely be assumed not to
depend on the phase $\psi $ at times before $t-\tau $, which determine $%
\widehat{\rho }_{gg}(t-\tau )$. The phase fluctuations are then described by
the independent-increment model \cite{Arnoldus}, which has the
phase-diffusion model as a special limiting case. The stochastic average of
the field correlation function then decays exponentially, according to the
equality $\left\langle \Omega ^{*}(t)\Omega (t-\tau )\right\rangle =\left|
\Omega \right| ^{2}\exp (-\mu \tau )$, with $\mu $ the bandwidth (halfwidth
at half maximum) of the Lorentzian profile. In this case the quantities $R$
and $L$ are determined by the equation
\begin{equation}
R+iL=\frac{\left| \Omega \right| ^{2}}{\mu +\gamma /2-i\delta }.
\label{stochphase}
\end{equation}
It is a simple check to notice that eqs. (\ref{popestoch}) and (\ref
{popgstoch}) reduce to the corresponding equations (\ref{pope}) and (\ref
{popg}) for monochromatic light when we substitute $\mu =0$. In that case we
simply find $R=\gamma S/2$ and $L=\delta S$.

In the case that the driving light also has intensity fluctuations, the
situation is more complex, and the factorization (\ref{stochav}) is not
exact. When the fluctuations are sufficiently weak and sufficiently rapid,
we can still assume the factorization as a reasonable approximation.
Therefore, the equations (\ref{popestoch}) and (\ref{popgstoch}) can be
assumed to be valid for broadband radiation in many situations of practical
interest. These equations, which strongly resemble the corresponding
equations (\ref{pope}) and (\ref{popg}) for monochromatic light, determine
the steady-state stochastic average of the density matrices $\widehat{\rho }%
_{ee}$ and $\widehat{\rho }_{gg}$. In the preceding sections we have
demonstrated that for monochromatic light and for all allowed values of $%
J_{e}$ and $J_{g}$ the steady-state density matrix obeys the commutations
rules (\ref{statementG}), so that the density matrix is diagonal in the
eigenstates of the light-shift operators. This means that the solutions of (%
\ref{pope}) and (\ref{popg}) do not depend at all on the strength of the
last terms in these equations. We conclude that the expressions for the
stochastically averaged steady-state solutions $\left\langle \widehat{\rho }%
_{ee}\right\rangle $ and $\left\langle \widehat{\rho }_{gg}\right\rangle $
in the absence of a dark state coincide with the solutions obtained in Sec.
V for $\widehat{\rho }_{ee}$ and $\widehat{\rho }_{gg}$, with the simple
replacement $S\rightarrow 2R/\gamma $. The structure of the dark states
follows from the defining equation (\ref{cptcond}), so that also the results
of Sec. IV are not modified by the fluctuations of the driving light. The
steady-state polarization properties of the atom are basically unaffected by
the finite bandwidth. The steady-state optical coherences $\widehat{\rho }%
_{eg}$ are best described by the expression for the stochastic average
\[
\left\langle \Omega ^{*}\widehat{\rho }_{eg}\right\rangle =(iR-L)\left[
\widehat{V}({\bf e})\left\langle \widehat{\rho }_{gg}\right\rangle
-\left\langle \widehat{\rho }_{ee}\right\rangle \widehat{V}({\bf e})\right]
,
\]
which follows immediately from eq. (\ref{stoch}). An expression containing $%
\widehat{\rho }_{ge}$ follows after Hermitian conjugation. Obviously, these
conclusions are valid exclusively when the light polarization displays no
fluctuations.

\section{Discussion and conclusions}

We have given a complete analytical and invariant description of the steady
state density matrix of a closed atomic dipole transition $J_{g}\to J_{e}$
driven by a resonant polarized radiation field. This is a long-standing
problem in atomic and optical physics. Solutions have been known for some
time in special cases of polarization and values of the angular momenta $J_e$
and $J_g$ of the excited and the ground state. The most complex class of
transitions occurs for $J_e = J_g +1$. In this case, the excited-state
density matrix can be highly non-anisotropic. It is remarkable, however,
that the anisotropy depends exclusively on the ellipticity of the
polarization, and it is unaffected by the frequency detuning of the
radiation from resonance, the light intensity or the spontaneous decay rate.
In the case that $J_e = J_g$ is half-integer, the excited-state density
matrix is fully isotropic in the steady state. In the remaining classes of
transitions ($J_e = J_g$ is integer, and $J_e = J_g -1$), the system has one
or two dark states, and the degree of excitation vanishes in the steady
state. For these cases, we give analytical invariant expressions for these
dark states for arbitrary elliptical polarization. These results are
interesting, not only from a fundamental point of view as an exact solution
of a quantum mechanical problem, but also since they can be used in numerous
applications.

As a first example, we mention the problem of non-linear propagation of
elliptically polarized light in a resonant gas medium. The steady-state
solution allows one to find the non-linear susceptibility tensor in
analytical form. The Doppler broadening is taken into account by the
substitution $\delta \to \delta -{\bf k}\cdot {\bf v}$ in the expressions
for $\widehat{\rho }_{eg}$, and then average over velocity.

A second case of interest is high-resolution polarization spectroscopy. The
Doppler-free resonances in the scheme of a strong pump and a weak probe
field can be directly evaluated by calculating the linear response to the
probe, in a steady state that is determined by the pump. Non-linear
interference effects between pump and probe are negligible in several cases,
e.g. when they are counter-propagating.

A third situation of practical importance occurs when cold atoms are slowly
moving through non-uniform radiation fields, with a position-dependent
amplitude $E_{0}({\bf r})$ and polarization vector ${\bf e}({\bf r})$. The
steady-state solution discussed in this paper can be viewed as the
zeroth-order approximation with respect to the atomic velocity. This
solution is needed for an explicit calculation of radiative forces \cite
{dalibard89,molmer94,taich96j,bezverbny98} and geometrical potentials \cite
{taich00,geompot}, which also affect the dynamics of atoms in optical
lattices.

Generally speaking, in many problems there are factors not taken into
account in our solution, such as finite interaction time, translational
motion of atoms, magnetic field etc. Very often these factors can be
considered as a small perturbation. In all these cases the steady-state
solution presented in this paper constitutes a zeroth-order approximation,
and thereby the first necessary step in the corresponding perturbation
treatment.

\acknowledgments

This work is partially supported by RFBR (grants \# 01-02-17036, and \#
01-02-17744), by a grant UR.01.01.062 of the Ministry of Education of the
Russian Federation, and by a grant INTAS-01-0855. It is also part of the
research program of the "Stichting voor Fundamenteel Onderzoek der Materie"
(FOM).

\appendix

\section{Spherical harmonics of a complex direction}

\label{spher_harm} \setcounter{equation}{0} \renewcommand{\theequation}{%
\thesection.\arabic{equation}}

Throughout the paper we use spherical harmonics that differ from the
standard definition \cite{varsh75} by a multiplicative factor
\[
n_{LM}=\sqrt{\frac{4\pi }{2L+1}}Y_{LM}\;.
\]
For arbitrary complex vector ${\bf a}={\bf a}^{\prime }+i{\bf a}^{\prime
\prime }$ a spherical harmonic of the rank $L$ is defined in terms of the
tensor constructions (\ref{aL}):
\begin{equation}
n_{LM}({\bf a})=\frac{1}{a^{L}}\sqrt{\frac{(2L-1)!!}{L!}}\{{\bf a}\}_{L}\;,
\label{Yfunc}
\end{equation}
where $a=\sqrt{({\bf a}\cdot {\bf a})}$. These generalized spherical
harmonics depend only on a direction in the complex three-dimensional space
\cite{manakov97}, i.e. they do not change under the transformation ${\bf a}%
\rightarrow \nu {\bf a}$ with $\nu $ an arbitrary complex number. For real
vectors (${\bf a}^{\prime \prime }=0$) the definition (\ref{Yfunc}) leads to
the standard spherical harmonics \cite{varsh75}. Starting from equation (\ref
{Yfunc}), one can derive the well-known formula \cite{varsh75}:
\begin{equation}
n_{LM}({\bf a})=e^{i\phi M}\sqrt{\frac{(L-M)!}{(L+M)!}}P_{L}^{M}(\cos \theta
)\;,  \label{extension}
\end{equation}
where $P_{L}^{M}(x)$ are the associated Legendre functions, and the complex
parameters $\theta $ and $\phi $ are expressed in terms of the spherical
components of vector ${\bf a}$ by
\[
\cos \theta =a_{0}/a\;,\;\;e^{2i\phi }=-a_{+1}/a_{-1}\,.
\]
Formula (\ref{extension}) can be regarded as a suitable analytic
continuation of the standard definition of the spherical harmonics $%
n_{LM}(\theta ,\phi )$ \cite{varsh75} to complex values of the angles $\phi $
and $\theta $ \cite{vilenkin65}. The definition (\ref{Yfunc}) is important
since the functions $n_{LM}$ obey the same group-theoretical relations as
ordinary spherical harmonics \cite{vilenkin65}. In particular, we indicate
the Clebsch-Gordan expansion of the product of two spherical harmonics of
the same argument
\begin{equation}
n_{l_{1}m_{1}}({\bf a})n_{l_{2}m_{2}}({\bf a})=\sum_{LM}C_{l_{1}0%
\;l_{2}0}^{L0}C_{l_{1}m_{1}\;l_{2}m_{2}}^{LM}n_{LM}({\bf a})
\label{cgexpansion}
\end{equation}
and the sum rule for the dot product of spherical harmonics of different
arguments \cite{vilenkin65,manakov97}
\begin{equation}
\left( n_{L}({\bf a})\cdot n_{L}({\bf b})\right) =P_{L}\left( \frac{({\bf a}%
\cdot {\bf b})}{ab}\right) \;\;,  \label{sumrule}
\end{equation}
where $P_{L}(x)$ are the Legendre polynomials.

\section{Calculating matrix elements}

\label{matr_el} \setcounter{equation}{0} \renewcommand{\theequation}{%
\thesection.\arabic{equation}}

The matrix elements of operators $\widehat{V}^{-1}$, $\left( \widehat{V}%
^{\dag }\widehat{V}\right) ^{-1}$, $\widehat{W}$, $\widehat{\widetilde{W}}$,
and $\widehat{X}$, which are used to write the steady-state density matrix $%
\widehat{\rho }$, can be determined in the natural coordinate frame. To be
specific, we fix the sign in eq. (\ref{conf2}):

\begin{equation}
{\bf e}=\sqrt{\cos (2\varepsilon )}{\bf e}_{0} - \sqrt{2}\sin (\varepsilon )%
{\bf e}_{+1}.  \label{expansion}
\end{equation}

\subsection{Transitions $J_g=J\rightarrow J_e=J$ with $J$ halfinteger}

In the natural coordinate frame the matrix $\widehat{V}$ is real and has a
lower triangular form with two nonzero diagonals:
\begin{equation}
\widehat{V}=\left(
\begin{array}{ccccccc}
\cdot &  &  &  &  &  & 0 \\
\cdot & \cdot &  &  &  &  &  \\
& \cdot & \cdot &  &  &  &  \\
&  & V_{\mu , \mu -1} & V_{\mu \mu } &  &  &  \\
&  &  & \cdot & \cdot &  &  \\
&  &  &  & \cdot & \cdot &  \\
0 &  &  &  &  & \cdot & \cdot
\end{array}
\right) ,  \label{vmatrix}
\end{equation}
where in accordance with the definitions (\ref{VQe}) and (\ref{expansion})
\begin{eqnarray}
V_{\mu \mu } &=&\frac{\mu }{\sqrt{J(J+1)}}\sqrt{\cos (2\varepsilon )}
\label{vdiag} \\
V_{\mu , \mu -1 } &=&-\sqrt{\frac{(J+\mu )(J-\mu +1)}{J(J+1)}}\,\sin
(\varepsilon ).
\end{eqnarray}
Its inverse matrix also is of the lower-triangular form and real. The matrix
elements of $\widehat{V}^{-1}$ are calculated by a direct method:
\begin{eqnarray}  \label{inv_v}
\left[ \widehat{V}^{-1}\right] _{\mu \mu ^{\prime }} &=&\frac{(-1)^{\mu -\mu
^{\prime }}}{V_{\mu ^{\prime }\mu ^{\prime }}}\prod_{\alpha =\mu ^{\prime
}+1}^{\mu }\frac{V_{\alpha (\alpha -1)}}{V_{\alpha \alpha }}=  \nonumber \\
&=&\sqrt{\frac{J(J+1)}{\cos (2\varepsilon )}}\left( \frac{\sin (\varepsilon )%
}{\sqrt{\cos (2\varepsilon )}}\right) ^{\mu -\mu ^{\prime }}\frac{1}{\mu
^{\prime }}\prod_{\alpha =\mu ^{\prime }+1}^{\mu }\frac{\sqrt{(J+\alpha
)(J-\alpha +1)}}{\alpha }.
\end{eqnarray}
The repeated products in (\ref{inv_v}) should be read while using the
conventions
\[
\prod_{\alpha =\mu +1}^{\mu }f_{\alpha }\equiv 1;\;\;\;\;\prod_{\alpha =\mu
^{\prime }+1}^{\mu }f_{\alpha }\equiv 0\;\;{\rm if }\;\;\mu ^{\prime }>\mu .
\]
Since the matrix $\widehat{V}$ is real, $\left( \widehat{V}^{\dagger
}\right) ^{-1}$ is obtained from $\widehat{V}^{-1}$ by transposition, i.e. $%
\left[ \left( \widehat{V}^{\dagger }\right) ^{-1}\right] _{\mu \mu ^{\prime
}}=\left[ \widehat{V}^{-1}\right] _{\mu ^{\prime }\mu }$. Thus, one can
easily write the matrix elements of $\left( \widehat{V}^{\dagger }\widehat{V}%
\right) ^{-1}$:
\begin{eqnarray}  \label{inv_vv}
&&\left[ \left( \widehat{V}^{\dagger }\widehat{V}\right) ^{-1}\right] _{\mu
\mu ^{\prime }}=(-1)^{\mu -\mu ^{\prime }}\sum_{\nu =-J}^{J}\frac{1}{V_{\nu
\nu }^{2}}\left( \prod_{\alpha =\nu +1}^{\mu }\frac{V_{\alpha (\alpha -1)}}{%
V_{\alpha \alpha }}\right) \left( \prod_{\alpha ^{\prime }=\nu +1}^{\mu
^{\prime }}\frac{V_{\alpha ^{\prime }(\alpha ^{\prime }-1)}}{V_{\alpha
^{\prime }\alpha ^{\prime }}}\right) =  \nonumber \\
&=&\frac{J(J+1)}{\cos (2\varepsilon )}\sum_{\nu =-J}^{J}\left( \frac{\sin
(\varepsilon )}{\sqrt{\cos (2\varepsilon )}}\right) ^{\mu +\mu ^{\prime
}-2\nu }\frac{1}{\nu ^{2}}\times  \nonumber \\
&&\times \left( \prod_{\alpha =\nu +1}^{\mu }\frac{\sqrt{(J+\alpha
)(J-\alpha +1)}}{\alpha }\right) \left( \prod_{\alpha ^{\prime }=\nu
+1}^{\mu ^{\prime }}\frac{\sqrt{(J+\alpha ^{\prime })(J-\alpha ^{\prime }+1)}%
}{\alpha ^{\prime }}\right) \;.
\end{eqnarray}

\subsection{Transitions $J_g=J\rightarrow J_e=J+1$}

In the natural coordinate frame the components of the spherical harmonics (%
\ref{Yfunc}) of the polarization vector ${\bf e}$ are written as
\begin{equation}  \label{Ypsi}
n_{L\;-M}({\bf e}) = (-1)^M\sqrt{\frac{(L+M)!}{(L-M)!}}\frac{1}{M!} \left(%
\frac{\sin \varepsilon}{\sqrt{\cos(2 \varepsilon)}}\right)^M \;,
\end{equation}
if $M \ge 0$, and $n_{L\,-M}({\bf e}) = 0$ for $M<0$. Substituting (\ref
{Ypsi}) into the definition (\ref{vlmatr}), we arrive at ($%
J_a=J+1,\;J_b=J,\;L=2J+1$):
\begin{eqnarray}  \label{wmatr}
W_{\mu m}&=&(-1)^{J-m}\frac{(2J+1+\mu-m)!}{(\mu-m)!} \sqrt{\frac{(2J+2)!(2J)!%
}{(4J+1)!(J+1+\mu)!(J+1-\mu)!(J+m)!(J-m)!}} \times  \nonumber \\
&\times& \left(\frac{\sin\varepsilon}{\sqrt{\cos(2\varepsilon)}}%
\right)^{\mu-m}\;,
\end{eqnarray}
where $\mu=-J-1, -J,\ldots J+1$, $m=-J, -J+1,\ldots J$ and $\mu-m \ge 0$.
The matrix $\widehat{\widetilde{W}}$ can be obtained from (\ref{wmatr})
using the time-reversal operation $\widetilde{W}_{m \mu}=(-1)^{J_g-m
-J_e-\mu}W_{-\mu\,-m}$.

In order to find matrix elements of $\widehat{X}$ we decompose it $\widehat{X%
}=\widehat{U}\widehat{W}$ using the Moore-Penrose pseudoinverse \cite
{mtheory} matrix $\widehat{U}$ with respect to $\widehat{V}$, i.e. $\widehat{%
V}\widehat{U}\widehat{V}=\widehat{V}$. The nonzero elements of $\widehat{V}$
are given by
\begin{eqnarray}  \label{vmatr}
V_{\mu ,\mu } &=&\sqrt{\frac{(J+1-\mu )(J+1+\mu )}{(J+1)(2J+1)}}\sqrt{\cos
(2\varepsilon )}  \nonumber \\
V_{\mu ,\mu -1} &=&\sqrt{\frac{(J+\mu )(J+1+\mu )}{(J+1)(2J+1)}}\sin
\varepsilon \;.
\end{eqnarray}
As the pseudoinverse matrix to $\widehat{V}$ we take the matrix with
elements
\begin{equation}
U_{m\mu }=\sqrt{\frac{(J+1)(2J+1)}{(J+1+\mu )(J+1-\mu )\cos (2\varepsilon )}}%
\left( -\frac{\sin \varepsilon }{\sqrt{\cos (2\varepsilon )}}\right) ^{m-\mu
}\prod_{\nu =\mu +1}^{m}\sqrt{\frac{J+\nu }{J+1-\nu }}\;,  \label{umatr}
\end{equation}
for $\mu =-J,-J+1,\ldots J$, supplemented by the zero columns $U_{m\;-J-1}=0$
and $U_{m\;J+1}=0$. The matrix multiplication of (\ref{umatr}) by (\ref
{wmatr}) yields the final result for the elements of $\widehat{X}$:
\begin{eqnarray}  \label{xmatr}
X_{m\,m^{\prime }} &=&\sqrt{\frac{(J+1)(2J+1)!(2J+2)!(J+m)!(J-m)!}{%
(4J+1)!(J+m^{\prime })!(J-m^{\prime })!\cos (2\varepsilon )}}\left( -\frac{%
\sin \varepsilon }{\sqrt{\cos (2\varepsilon )}}\right) ^{m-m^{\prime }}\times
\nonumber \\
&\times &\sum_{\mu =m^{\prime }}^{m}(-1)^{J-\mu }\frac{(2J+1+\mu -m^{\prime
})!}{(\mu -m^{\prime })!(J+1+\mu )!(J+1-\mu )!}\;.
\end{eqnarray}

\section{Algebra of the operators $\widehat{V}^{ab}_{L}({\bf a})$}

\label{alg_vl} \setcounter{equation}{0} \renewcommand{\theequation}{%
\thesection.\arabic{equation}}

Using the standard Racah algebra \cite{racah}, one can write a general
expression for products of the operators $\widehat{V}$ (\ref{vlmatr}) with
different ranks:
\begin{equation}  \label{algebra}
\widehat{V}_{L_1}^{ab}({\bf a})\widehat{V}_{L_2}^{bc}({\bf b}) = \sum_{K}
(-1)^{J_a+J_c+L_1+L_2} \Pi_{L_1,L_2} \left\{
\begin{array}{rcl}
K & L_1 & L_2 \\
J_b & J_c & J_a
\end{array}
\right\} \left( \left\{ n_{L_1}({\bf a})\otimes n_{L_2}({\bf b})\right\}_{K}
\cdot \widehat{T}^{ac}_{K} \right) \,,
\end{equation}
where $\Pi_{x,y,\ldots}=\sqrt{(2x+1)(2y+1)\ldots}\;$ is the standard
notation of the handbook \cite{varsh75}. In the special case that ${\bf b}=%
{\bf a}$, after using eq. (\ref{cgexpansion}) we obtain from (\ref{algebra})
an analogue of the Clebsch-Gordan expansion:
\begin{equation}  \label{vl_cge}
\widehat{V}_{L_1}^{ab}({\bf a})\widehat{V}_{L_2}^{bc}({\bf a}) = \sum_{K}
(-1)^{J_a+J_c+L_1+L_2} \Pi_{L_1,L_2} C^{K0}_{L_1 0\;L_2 0} \left\{
\begin{array}{rcl}
K & L_1 & L_2 \\
J_b & J_c & J_a
\end{array}
\right\} \widehat{V}_{K}^{ac}({\bf a}) \;.
\end{equation}
Equations (\ref{algebra}) and (\ref{vl_cge}) lead to the following
relationships: \newline
1) For arbitrary ranks $L_1$ and $L_2$, and for arbitrary angular momenta $%
J_{a}$ and $J_{b}$ we find
\begin{equation}  \label{first}
\widehat{V}^{ab}_{L_1}({\bf a})\widehat{V}_{L_2}^{ba}({\bf a}) = \widehat{V}%
^{ab}_{L_2}({\bf a})\widehat{V}_{L_1}^{ba}({\bf a}) \;,
\end{equation}
since both sides have the same expansion in the tensor operators $\widehat{V}%
^{aa}_{K}({\bf a})$. Here we use the symmetry of (\ref{vl_cge}) with respect
to the permutation $L_1 \leftrightarrow L_2$ at $J_a=J_c$. \newline
2) Depending on the class of transition, for arbitrary vectors ${\bf a}$ and
${\bf b}$ we find:

a) for transitions $J_g=J \rightarrow J_e=J$
\begin{equation}  \label{second1}
\widehat{V}_1^{ge}({\bf b})\widehat{V}_0^{eg}({\bf a}) = \widehat{V}_0^{ge}(%
{\bf a})\widehat{V}_1^{eg}({\bf b}) \;.
\end{equation}

b) for transitions $J_g=J \rightarrow J_e=J+1$
\begin{equation}  \label{second2}
\widehat{V}_1^{ge}({\bf a})\widehat{V}_{2J+1}^{eg}({\bf b}) = \widehat{V}%
_{2J+1}^{ge}({\bf b})\widehat{V}_1^{eg}({\bf a}) \;.
\end{equation}

c) for transitions $J_{g}=J\rightarrow J_{e}=J-1$
\begin{equation}
\widehat{V}_{1}^{eg}({\bf a})\widehat{V}_{2J-1}^{ge}({\bf b})=\widehat{V}%
_{2J-1}^{eg}({\bf b})\widehat{V}_{1}^{ge}({\bf a})\;.  \label{second3}
\end{equation}
The property (\ref{second1}) is obvious, if we recall that in this case the
operator $\widehat{V}_{0}^{ge}$ is proportional to the unit matrix and $%
\widehat{V}_{1}^{ge}({\bf b})=\widehat{V}_{1}^{eg}({\bf b})$. To prove the
validity of (\ref{second2}) it is sufficient to expand both sides of (\ref
{second2}) in the operators $\widehat{T}_{Kq}^{gg}$ and allow for the fact
that all ranks except $K=2J$ are forbidden by the selection rules contained
in the $6j$ symbols in (\ref{algebra}). The equation (\ref{second2}) then
reduces to the identity $\{n_{1}({\bf a})\otimes n_{2J+1}({\bf b}%
)\}_{2J}\equiv \{n_{2J+1}({\bf b})\otimes n_{1}({\bf a})\}_{2J}$, which
holds since the number $(2J+1)+1-2J=2$ is even. The equation (\ref{second3})
can be proved in a similar way.

\begin{figure}[tbp]
\begin{center}
\mbox{\psfig{file=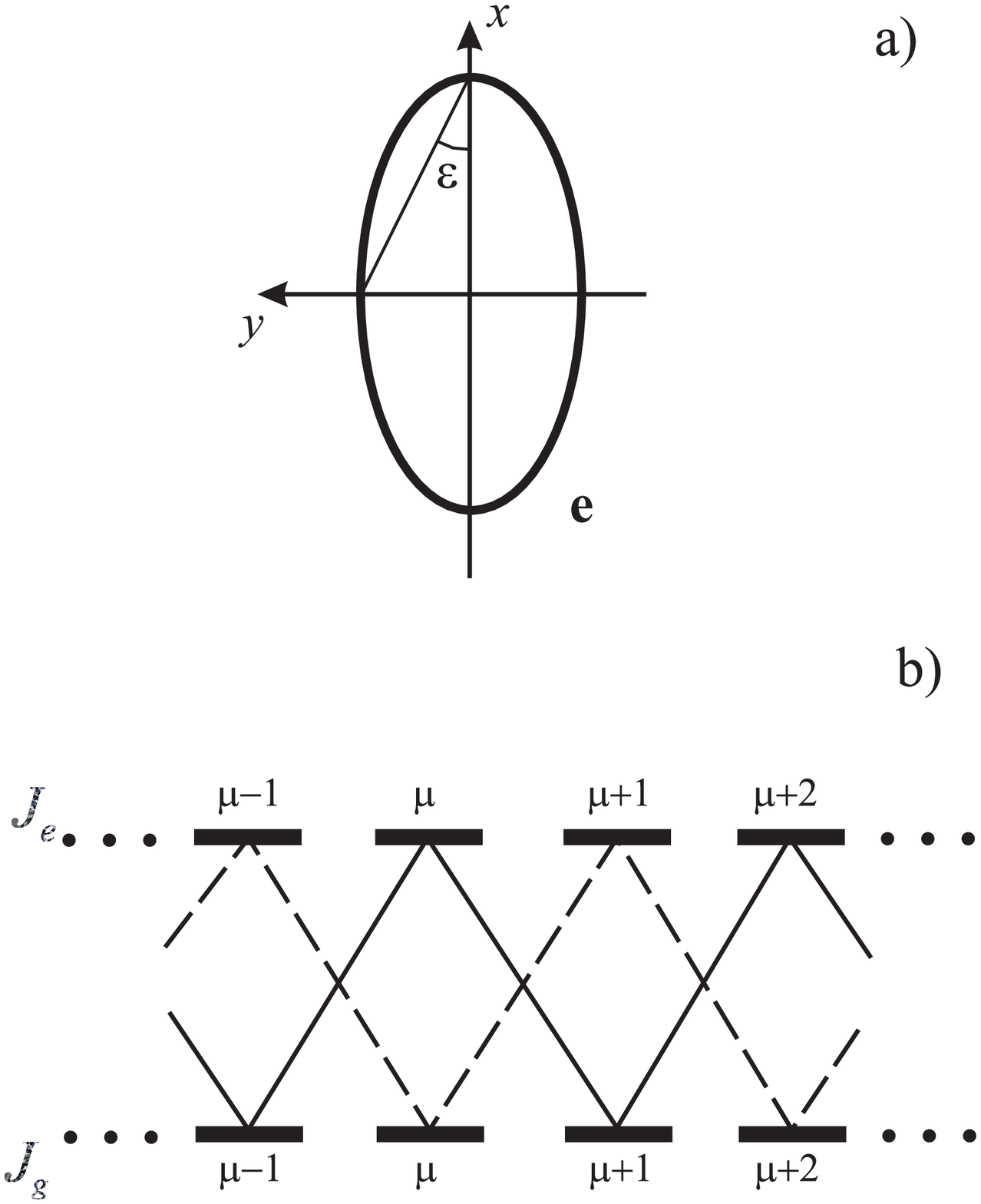,width=4.5in}}
\end{center}
\caption{Conventional coordinate frame for the representation of elliptical
polarization. a) polarization ellipse lies in $xy$-plane, with major axis in
$x$-direction. b) transition scheme with the $z$-axis as quantization axis.}
\end{figure}

\begin{figure}[tbp]
\begin{center}
\mbox{\psfig{file=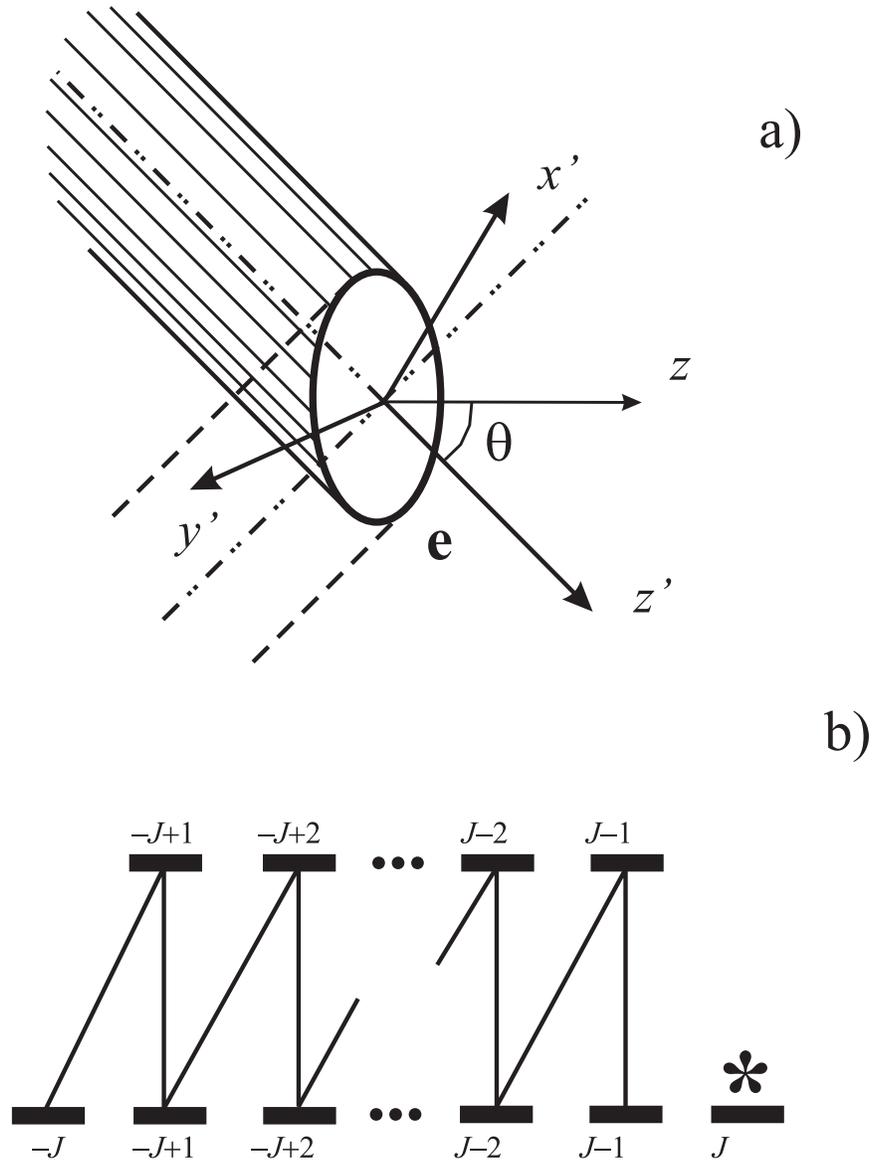,width=4.5in}}
\end{center}
\caption{Natural coordinate frame for the representation of elliptical
polarization. a) polarization vector is superposition of circular
polarization defined by cylinder (in $x^{\prime }y^{\prime }$-plane), and
linear polarization along axis of cylinder ($z^{\prime }$-axis). b)
transition scheme with the $z^{\prime }$-axis as quantization axis.}
\end{figure}

\begin{figure}[tbp]
\begin{center}
\mbox{\psfig{file=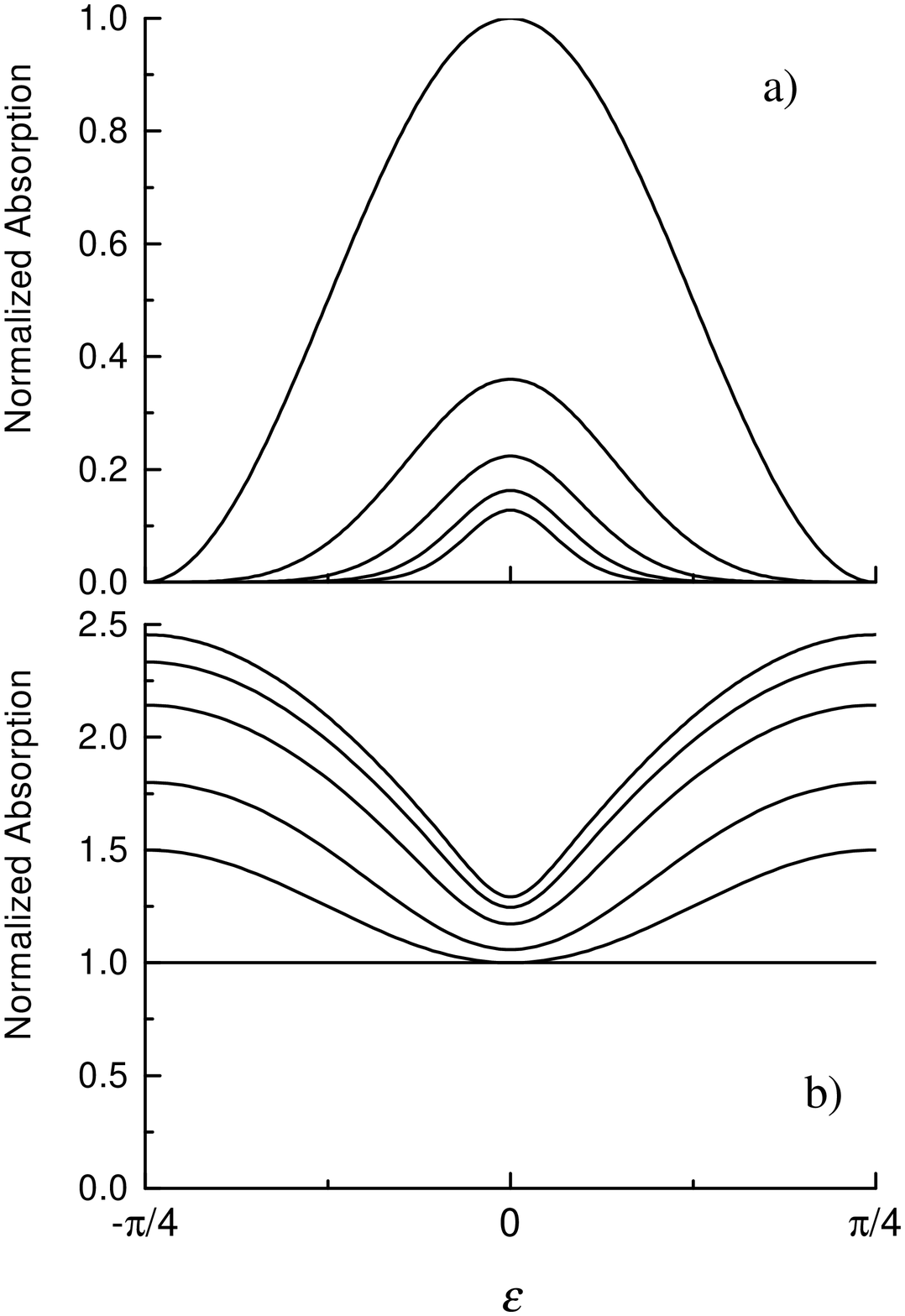,width=4.5in}}
\end{center}
\caption{Total absorption at low saturation versus the ellipticity $%
\varepsilon $. (a) transitions $J\to J$ with $J$ a half-integer; $J$ runs
from $1/2$ to $9/2$ (from top to bottom). (b) transitions $J\to J+1$; $%
J=0,\,1/2,\,1,\,2,\,3,\,4$ (from bottom to top). All curves are normalized
to the linear absorption. }
\label{fig3}
\end{figure}


\begin{references}
\bibitem{eberly78}  L. Allen and J. H. Eberly {\em Optical Resonance and
Two-Level Atoms} (John Wiley \& Sons, New York-London-Sydney-Toronto, 1975).

\bibitem{let&cheb77}  V. S. Letokhov and V. P. Chebotaev, {\em Principles of
Nonlinear Laser Spectroscopy}, (Springer, Berlin, 1977).

\bibitem{rautian79}  S. G. Rautian and A. M. Shalagin, {\em Kinetic Problems
of Nonlinear Spectroscopy} (Elsevier, Amsterdam, 1991).

\bibitem{minogin87}  V. G. Minogin and V. S. Letokhov, {\em Laser Light
Pressure on Atoms} (Gordon and Breach, New York, 1987).

\bibitem{kazantsev91}  A. P. Kazantsev, G. I. Surdutovich, V. P. Yakovlev,
{\em Mechanical Action of Light on Atoms} (World Scientific, Singapore,
1990).

\bibitem{smirnov89}  V. S. Smirnov, A. M. Tumaikin, V. I. Yudin, JETP {\bf 69%
}, 913 (1989).

\bibitem{lasercooling}  Special Issue {\em Laser Cooling and Trapping of
Atoms}, J. Opt. Soc. Am. B {\bf 6}, 11 (1989); Special Issue {\em Laser
Cooling and Trapping}, Laser Physics {\bf 4}, 5 (1994); S. Chu, Rev. Mod.
Phys. {\bf 70}, 685 (1998); C. Cohen-Tannoudji, Rev. Mod. Phys. {\bf 70},
707 (1998); W. Phillips, Rev. Mod. Phys. {\bf 70}, 721 (1998).

\bibitem{manikin84}  E. A. Manykin and V. V. Samartsev, {\em Optical
Echo-Spectroscopy} [in Russian] (Nauka, Moscow, 1984).

\bibitem{PPE}  N. N. Rubtsova, L. S. Vasilenko, E. B. Khvorostov, Laser
Physics {\bf 9}, 239 (1999), and references cited therein.

\bibitem{velich89}  A. M. Akulshin, V. L. Velichansky, M. V. Krasheninnikov,
V. S. Smirnov, A. M. Tumaikin, V. I. Yudin, Zh. Eksp. Teor. Fiz. {\bf 96},
107 (1989) [JETP {\bf 69}, 58 (1989)].

\bibitem{Dalibard}  J. Dalibard and C. Cohen-Tannoudji, J. Opt. Soc. Amer. B
{\bf 6}, 2023 (1989).

\bibitem{sctheory}  Y. Castin and K. M\o lmer, J. Phys. B {\bf 23}, 4101
(1990); K. M\o lmer, Phys. Rev. A {\bf 44}, 5820 (1991); J. Javanainen,
Phys. Rev. A {\bf 44}, 5857 (1991); S.M. Yoo and J. Javanainen, Phys. Rev. A
{\bf 45}, 3071 (1992); V. Finkelstein, P.R. Berman, J. Guo, Phys. Rev. A
{\bf 45}, 1829 (1992); J. Werner, H. Wallis, G. Hillenbrand, A. Steane, J.
Phys. B {\bf 26}, 3063 (1993); S. Chang, T. Y. Kwon, H. S. Lee, V. Minogin,
Phys. Rev. A {\bf 60}, 2308 (1999); S. Chang, T. Y. Kwon, H. S. Lee, V.G.
Minogin, Phys. Rev. A {\bf 64}, 013404 (2001).

\bibitem{berman91}  P.R. Berman, Phys. Rev. A {\bf 43}, 1470 (1991); P.R.
Berman, G. Rogers, and B. Dubetsky, Phys. Rev. A {\bf 48}, 1506 (1993).

\bibitem{nienhuis91}  G. Nienhuis, P. van der Straten, S.-Q. Shang, Phys.
Rev. A {\bf 44}, 462 (1991).

\bibitem{macek74}  J. Macek and I. Hertel, J. Phys. B {\bf 7}, 2173 (1974).

\bibitem{nienhuis82}  G. Nienhuis, Phys. Rev. A {\bf 26}, 3137 (1982).

\bibitem{gao93}  Bo Gao, Phys. Rev. A {\bf 48}, 2443 (1993).

\bibitem{dalibard84}  J. Dalibard, S. Reynaud and C. Cohen-Tannoudji, J.
Phys. B {\bf 17}, 4577 (1984).

\bibitem{dalibard89}  J. Dalibard and C. Cohen-Tannoudji, J. Opt. Soc. Am. B
{\bf 6}, 2023 (1989).

\bibitem{suter91}  D. Suter, Opt. Commun. {\bf 86}, 381 (1991).

\bibitem{davis92}  W. D. Davis, A. L. Gaeta and R. W. Boyd, Optics Lett.
{\bf 17}, 1304 (1992).

\bibitem{molmer94}  K. M\o lmer and C. Westbrook, Laser Physics {\bf 4}, 872
(1994).

\bibitem{taich95}  A. V. Taichenachev, A. M. Tumaikin, V. I. Yudin, G.
Nienhuis, Zh. Eksp. Teor. Fiz. {\bf 108}, 415 (1995) [JETP {\bf 81}, 224
(1995)].

\bibitem{taich99epl}  A. V. Taichenachev, A. M. Tumaikin, V. I. Yudin,
Europhys. Lett. {\bf 45}, 301 (1999).

\bibitem{nienhuis98epl}  G. Nienhuis, A. V. Taichenachev, A. M. Tumaikin, V.
I. Yudin, Europhys. Lett. {\bf 44}, 20 (1998).

\bibitem{bezverbny00}  A.V. Bezverbny, Zh. Eksp. Teor. Fiz. {\bf 118}, 1066
(2000) [JETP {\bf 91}, 921 (2000)].

\bibitem{nasyrov01}  K.A. Nasyrov, Phys. Rev. A {\bf 63}, 043406 (2001).

\bibitem{Aspect}  A. Aspect, E. Arimondo, R. Kaiser, N. Vansteenkiste, and
C. Cohen-Tannoudji, Phys. Rev. Lett. {\bf 61}, 826 (1988); A. Aspect, E.
Arimondo, R. Kaiser, N. Vansteenkiste, and C. Cohen-Tannoudji, J. Opt. Soc.
Amer. B {\bf 6}, 2112 (1989).

\bibitem{vscpt}  F. Mauri and E. Arimondo, Europhys. Lett. {\bf 16}, 717
(1991).; M. A. Olshanii, J. Phys. B {\bf 24}, L583 (1991); A.V.
Taichenachev, A.M. Tumaikin, V.I. Yudin, M. A. Olshanyi, Pis'ma v Zh. Eksp.
Teor. Fiz. {\bf 53}, 336 (1991) [JETP Lett. {\bf 53}, 351 (1991)]; F. Mauri
and E. Arimondo, Appl. Phys. B {\bf 54}, 420 (1992); E. Papoff, F. Mauri, E.
Arimondo, J. Opt. Soc. Am. B {\bf 9}, 321 (1992); R. Gupta, S. Padua, C.
Xie, H. Batelaan, and H. Metcalf, J. Opt. Soc. Amer. B {\bf 11}, 537 (1994);
G. Morigi, B. Zambon, N. Leinfellner, E. Arimondo, Phys. Rev. A {\bf 53},
2616 (1996); C. Menotti, P. Horak, H. Ritsch, J. H. M\"{u}ller, E. Arimondo,
Phys. Rev. A {\bf 56}, 2123 (1997); C. Menotti, G. Morigi, J. H. M\"{u}ller,
E. Arimondo, Phys. Rev. A {\bf 56}, 4327 (1997).

\bibitem{invvscpt}  M.A. Olshanii and V.G. Minogin, Opt. Commun. {\bf 89,}
393 (1992); J. Lawall, F. Bardou, B. Saubamea, K. Shimizu, M. Leduc, A.
Aspect, and C. Cohen-Tannoudji, Phys. Rev. Lett.{\bf 73}, 1915 (1994); J.
Lawall, S. Kulin, B. Saubamea, N. Bigelow, M. Leduc, C. Cohen-Tannoudji,
Phys. Rev. Lett. {\bf 75}, 4194 (1995).

\bibitem{tumaikin90}  A. M. Tumaikin and V. I. Yudin, Zh. Eksp. Teor. Fiz.
{\bf 98}, 81 (1990) [JETP {\bf 71}, 43 (1990)].

\bibitem{born}  M. Born and E. Wolf, {\em Principles of Optics} (Pergamon,
London, 1959).

\bibitem{nienhuis86}  G. Nienhuis, Opt. Commun. {\bf 59}, 353 (1986).

\bibitem{taich98sp}  A. V. Taichenachev, A. M. Tumaikin, V. I. Yudin, G.
Nienhuis, Zh. Eksp. Teor. Fiz. {\bf 114}, 125 (1998) [JETP {\bf 87}, 70
(1998)].

\bibitem{varsh75}  D. A. Varshalovich, A. N. Moskalev, V. K. Khersonsky,
{\em Quantum Theory of Angular Momentum} (World Scientific, Singapore, 1988).

\bibitem{manakov96}  N. L. Manakov, S. I. Marmo, and A. V. Meremianin, J.
Phys. B: At. Mol. Opt. Phys. {\bf 29}, 2711, (1996).

\bibitem{manakov97}  N. L. Manakov and A. V. Merem'yanin, Zh. Eksp. Teor.
Fiz. {\bf 111}, 1984 (1997) [JETP {\bf 84}, 1080 (1997)].

\bibitem{vilenkin65}  N. Ya. Vilenkin, {\em Special Functions and the Theory
of Group Representations} (American Mathematical Society, Providence, 1968).

\bibitem{mtheory}  F. R. Gantmacher, {\em The Theory of Matrices} (Chelsea,
New York, 1959).

\bibitem{racah}  U. Fano and G. Racah, {\em Irreducible Tensorial Sets}
(Academic, New York, 1959).

\bibitem{taich96j}  A. V. Taichenachev, A. M. Tumaikin, V. I. Yudin, Zh.
Eksp. Teor. Fiz. {\bf 110}, 1727 (1996) [JETP {\bf 83}, 949 (1996)].

\bibitem{taich00}  A. V. Taichenachev, A. M. Tumaikin, V. I. Yudin, Zh.
Eksp. Teor. Fiz. {\bf 118}, 77 (2000). [JETP {\bf 91}, 67 (2000)].

\bibitem{taich96jl}  A. V. Taichenachev, A. M. Tumaikin, V. I. Yudin, Pis'ma
v Zh. Eksp. Teor. Fiz. {\bf 64}, 8 (1996) [JETP Lett. {\bf 64}, 7 (1996)].

\bibitem{EIA}  A. M. Akulshin, S. Barreiro, A. Lezama, Phys. Rev. A {\bf 57}%
, 2996 (1998); A. Lezama, S. Barreiro, A. M. Akulshin, Phys. Rev. A {\bf 59}%
, 4732 (1999); A. V. Taichenachev, A. M. Tumaikin, V. I. Yudin,
Phys. Rev. A {\bf 61}, 011802 (2000); Y. Dancheva, G. Alzetta, S.
Cartaleva, M. Taslakov, Ch. Andreva, Optics Comm. {\bf 178}, 103
(2000); F. Renzoni, S. Cartaleva, G. Alzetta, E. Arimondo, Phys.
Rev. A {\bf 63}, 065401 (2001); A. V. Papoyan, M. Auzinsh, K.
Bergmann, Eur. Phys. J. D {\bf 21}, 63 (2002); C. Goren, A. D.
Wilson-Gordon, M. Rosenbluh, H. Friedmann, Phys. Rev. A {\bf 67},
033807 (2003); H. Failache, P. Valente, G. Ban, V. Lorent, A. Lezama,
Phys. Rev. A {\bf 67}, 043810 (2003).

\bibitem{Arnoldus}  H.F. Arnoldus and G. Nienhuis, J. Phys. B: At. Mol.
Phys. {\bf 16}, 2325 (1983).

\bibitem{bezverbny98}  A. V. Bezverbnyi, A. M. Tumaikin, G. Nienhuis, Opt.
Commun. {\bf 148}, 151 (1998).

\bibitem{geompot}  A. V. Taichenachev, A. M. Tumaikin, V. I. Yudin, Laser
Physics {\bf 2,} 575 (1992); R. Dum and M. Olshanii, Phys. Rev. Lett. {\bf 76%
}, 1788 (1996); P. M. Visser and G. Nienhuis, Phys. Rev. A {\bf 57}, 4581
(1998).
\end{references}
\end{document}